\documentclass[longauth]{aa}
\usepackage{natbib}
\usepackage{float}
\usepackage{xcolor}
\bibpunct{(}{)}{;}{a}{}{,}
\usepackage{txfonts}%
\usepackage[T1]{fontenc}%
 
\usepackage{graphicx}
\usepackage{xspace}
\usepackage{hyperref}
\hypersetup{hidelinks}
\usepackage{array}
\usepackage{arydshln}
\usepackage{amsmath}
\usepackage{amssymb}%

\newcommand{\hrieuv}{HRI\textsubscript{EUV}\xspace}
\newcommand{\hrilya}{HRI\textsubscript{Lya}\xspace}

\begin{document}

\title{Height-dependent thermal properties of extreme-ultraviolet campfires in the quiet Sun}
\titlerunning{Height-dependent thermal properties of EUV campfires in the quiet Sun}

\author{O.~Podladchikova\inst{\ref{i:aip},\ref{i:pmod}}\fnmsep\thanks{Corresponding author: O.~Podladchikova, \email{epodlad@gmail.com}}
         \and
        A.~Warmuth\inst{\ref{i:aip}}
           \and 
        L.~Harra\inst{\ref{i:pmod},\ref{i:eth}}
           \and  
        C.~Verbeeck\inst{\ref{i:rob}}
           \and     
        M.~K.~Georgoulis\inst{\ref{i:apl},\ref{i:acad_at}}
           \and                      
        A.~M.~Veronig \inst{\ref{i:graz},\ref{i:kanz}}
           \and   
        D.~M\"uller\inst{\ref{i:esa1}}
             \and   
      S.~J.~Hofmeister\inst{\ref{i:columbia},\ref{i:aip}}
           \and   
        P.~Antolin\inst{\ref{i:north}}
           \and      
        M.~Haberreiter\inst{\ref{i:pmod}}
           \and      
        S.~Purkhart\inst{\ref{i:graz}}
            \and   
        D.~M.~Long\inst{\ref{i:dcu}}
           \and   
        A.~F.~Battaglia\inst{\ref{i:irsol}}   
            \and   
        N.~Engler\inst{\ref{i:pmod},\ref{i:eth}}
            \and    
        É.~Buchlin\inst{\ref{i:ias}}
           \and          
        L.~Dolla\inst{\ref{i:rob}}
           \and  
        M.~Mierla\inst{\ref{i:rob},\ref{i:igar}}
             \and  
        L.~Rodriguez\inst{\ref{i:rob}}
            \and        
        M.~J.~West\inst{\ref{i:esa1}}
            \and                     
        A.~N.~Zhukov\inst{\ref{i:rob}}%
            \and            
        E.~Soubrié\inst{\ref{i:ias},\ref{i:iaccc}}
             \and
        V.~B\"{u}chel\inst{\ref{i:pmod}}
             \and
        A.~De~Groof\inst{\ref{i:esac}}
             \and
        M.~Gyo\inst{\ref{i:pmod}}
             \and
        J.~P.~Halain\inst{\ref{i:esa1},\ref{i:csl}}
            \and
        B.~Inhester\inst{\ref{i:mps}}
             \and
        E.~Kraaikamp\inst{\ref{i:rob}}
             \and
        D.~Pfiffner\inst{\ref{i:pmod}}
             \and       
        P.~Rochus$^\dagger$\inst{\ref{i:csl}}\fnmsep\thanks{In memoriam}
             \and
        F.~Schuller\inst{\ref{i:aip}}
            \and
        P.~J.~Smith\inst{\ref{i:mssl}}
            \and
        W.~Schmutz\inst{\ref{i:pmod}}
             \and
        K.~Stegen\inst{\ref{i:rob}}
}

\institute{
            Leibniz Institute for Astrophysics Potsdam,
            An der Sternwarte 16,
            14482 Potsdam, Germany \label{i:aip}
            \and
            Physikalisch-Meteorologisches Observatorium Davos, World Radiation Center,
            7260 Davos Dorf, Switzerland \label{i:pmod}
            \and
            ETH-Z\"urich, Wolfgang-Pauli-Str. 27, 8093 Z\"urich, Switzerland \label{i:eth}
            \and
            Solar-Terrestrial Centre of Excellence -- SIDC, Royal Observatory of Belgium,
            Ringlaan -3- Av. Circulaire, 1180 Brussels, Belgium \label{i:rob}
            \and
            Johns Hopkins University Applied Physics Laboratory,
            11100 Johns Hopkins Rd, Laurel, MD 20723, USA \label{i:apl}
            \and
            Research Center for Astronomy and Applied Mathematics (RCAAM), Academy of Athens,
            11527 Athens, Greece (on leave) \label{i:acad_at}
            \and
            University of Graz, Institute of Physics,
            Universit\"atsplatz 5, 8010 Graz, Austria \label{i:graz}
            \and
            University of Graz, Kanzelh\"ohe Observatory for Solar and Environmental Research,
            9521 Treffen, Austria \label{i:kanz}
            \and
            European Space Agency (ESA/ESTEC), 2200 AG Noordwijk, The Netherlands \label{i:esa1}
            \and
            Columbia Astrophysics Laboratory, Columbia University,
            550 West 120th Street, New York, NY 10027, USA \label{i:columbia}
            \and
            School of Engineering, Physics and Mathematics,
            Northumbria University, Newcastle Upon Tyne, NE1 8ST, UK \label{i:north}
            \and
            Centre for Astrophysics \& Relativity, School of Physical Sciences, Dublin City 
            University, Glasnevin Campus, Dublin, D09 V209, Ireland \label{i:dcu}
            \and
            IRSOL, Istituto Ricerche Solari Aldo e Cele Daccò,
            Universit\`{a} della Svizzera italiana,
            6605 Locarno Monti, Switzerland \label{i:irsol}
            \and
            Institute of Geodynamics, Romanian Academy, 020032 Bucharest, Romania \label{i:igar}
            \and
            Universit\'{e} Paris-Saclay, CNRS, Institut d'Astrophysique Spatiale,
            91405 Orsay, France \label{i:ias}
            \and
            Institute of Applied Computing \& Community Code,
            Universitat de les Illes Balears, 07122 Palma de Mallorca, Spain \label{i:iaccc}
            \and
            European Space Agency (ESA/ESAC), E-28692 Villanueva de la 
            Ca\~nada, Madrid, Spain \label{i:esac}
            \and
            Centre Spatial de Li\`ege, Universit\'e de Li\`ege,
            4031 Angleur, Belgium \label{i:csl}
            \and
            Max Planck Institute for Solar System Research,
            37077 G\"ottingen, Germany \label{i:mps}
            \and
            UCL-Mullard Space Science Laboratory,
            Holmbury St. Mary, Dorking, Surrey RH5 6NT, UK \label{i:mssl}
}

\date{Received ...; accepted...}

\abstract
{Small-scale impulsive energy-release events are widely considered 
a key ingredient of quiet-Sun coronal heating. The discovery of 
extreme ultraviolet (EUV) campfires by Solar Orbiter has opened a new 
observational window at unprecedented spatial scales.}
{We quantified the thermal-energy content of EUV campfires, examined
their impulsive heating characteristics, and investigated how their 
properties vary with atmospheric height.}
{We analysed the complete set of campfires obtained in a single 
observing sequence on 30~May~2020 
with Solar Orbiter/EUI HRI$_{\rm EUV}$ at 174~\AA. Temperatures 
and emission measures were adopted from SDO/AIA diagnostics, and 
geometric uncertainties were explored using multiple volume models. 
Statistical properties were characterised using cumulative 
distribution functions and maximum-likelihood techniques, and 
results were compared with Solar Orbiter/STIX microflares and 
SDO/AIA nanoflare-like events for context.}
{Campfire durations show a clear dependence on height: short-duration events occur at all heights, whereas 
longer-lived brightenings are preferentially associated with
chromospheric layers. Emitting volumes and thermal energies, by 
contrast, show no comparable height dependence. 
The background-subtracted thermal energies span 
$\sim10^{20}$--$10^{24}$~erg, with rare events reaching 
$\sim5\times10^{24}$~erg. The statistically robust power-law 
regime extends over $\sim10^{22}$--$10^{24}$~erg, while 
non-background-subtracted estimates are systematically higher 
by approximately one order of magnitude. Directly detected 
campfires contribute a sub-percent fraction of the canonical 
quiet-Sun heating requirement. Accounting approximately for the 
detection threshold and differential emission measure completeness 
could raise this contribution to a value of order a few percent, though 
this remains a single-epoch, order-of-magnitude estimate.}
{Extreme ultraviolet  campfires extend the flare energy distribution towards lower 
energies and provide height-resolved constraints on impulsive 
heating in the quiet Sun. The height dependence of event 
durations -- driven mainly by an excess of long-lived events in 
the lower chromosphere -- combined with broadly invariant energy 
and volume distributions, indicates that atmospheric stratification 
primarily modulates cooling and dissipation timescales rather 
than the characteristic energy scale of the heating process.}
\maketitle

\section{Introduction}

Understanding the mechanisms responsible for heating the solar corona 
remains one of the central open questions of solar physics. Maintaining 
coronal temperatures requires a continuous energy input from the solar 
magnetic field, yet the physical processes by which this energy is 
transported and dissipated remain debated. Both observations and theory 
indicate that coronal heating is fundamentally impulsive, with energy 
released through numerous small-scale events -- driven by magnetic 
reconnection, turbulent wave dissipation, or both -- rather than as a 
steady process.

Coronal heating is now understood as a multi-scale problem involving 
the generation, transport, and dissipation of magnetic energy across a 
broad range of spatial and temporal scales. Two broad paradigms have 
traditionally been invoked: wave-based heating and impulsive 
(nanoflare-like) heating. While often treated as distinct, these 
processes represent complementary manifestations of the same underlying 
physics -- the conversion of magnetic energy into plasma heating through 
electric currents and electromagnetic stresses \citep{Benz2008, 
Klimchuk2015}. Photospheric motions inevitably stress coronal magnetic 
fields, generating currents throughout the atmosphere that dissipate 
either directly or through wave--particle interactions, depending on 
plasma conditions and magnetic topology.

A useful physical distinction can be drawn between dissipation driven by 
currents perpendicular to the magnetic field and those aligned parallel 
to it \citep{SpicerBrown1980, PriestForbes2000}. Perpendicular currents 
drive magnetic reconnection on Alfv\'enic timescales through extremely 
thin current sheets, whereas field-aligned current dissipation proceeds 
through anomalous resistivity and current-driven instabilities, 
converting magnetic energy more directly into heat. In both cases, 
energy release occurs in collisionless plasma through wave--particle 
interactions \citep{Leontovich1975}. In the low corona and upper 
transition region, where plasma-$\beta$ approaches unity, quasi-parallel 
current sheets are expected to be systematically wider than those forming 
higher in the corona, favouring gentler, thermally dominated energy 
release \citep{Demoulin2000}.

Wave heating and impulsive heating should therefore be viewed not as 
competing mechanisms but as limiting cases of magnetic-energy 
dissipation operating on different scales. Wave dissipation is 
particularly effective in long coronal loops and open magnetic 
structures, where Alfv\'en waves cascade to small scales via turbulence 
\citep{Heyvaerts1983}. Observational evidence shows that Alfv\'enic 
waves also dissipate at low coronal heights, contributing to the heating of 
short loops and quiet-Sun regions \citep[e.g.][]{Tomczyk2007, 
Hahn2014, Hahn2025}. The two scenarios have a common requirement: magnetic 
energy injected at large scales must cascade efficiently to the small 
scales where dissipation occurs.

Extreme-ultraviolet (EUV) and X-ray diagnostics measure macroscopic plasma properties -- 
temperature and emission measure -- that represent statistical averages 
over many unresolved dissipation sites. As noted by \citet{Benz2008}, 
these thermodynamic quantities are fully consistent with kinetic 
descriptions of energy release even without direct evidence of 
non-thermal particles. Magnetohydrodynamic (MHD) descriptions capture the large-scale plasma 
response, while the actual energy conversion occurs at much smaller, 
often unresolvable scales.

A broad theoretical consensus holds that the corona is heated by a 
hierarchy of impulsive events that span a wide energy range. The nanoflare 
hypothesis \citep{Parker1988} formalised this picture, and subsequent 
simulations and statistical studies have shown that such heating naturally 
produces scale-free energy distributions and intermittent plasma response 
\citep[e.g.][]{Vlahos1995, Georgoulis1998, Rappazzo2008, Guerreiro2017}. 
\citet{Klimchuk2015,Klimchuk2017} emphasised that nanoflares are 
expected to remain below direct observational resolution, with EUV and 
X-ray observations capturing only their macroscopic thermal signatures. 
Wave-heating models -- particularly those involving counter-propagating 
Alfv\'en waves -- remain equally viable and have received strong 
observational support \citep[e.g.][]{Tomczyk2007, Hahn2014, 
DeMoortel2015, Cranmer2017, Hahn2025}. The two paradigms are not 
mutually exclusive and may operate together or in different regions.

Recent instrumental advances have begun to probe this regime directly. 
High-resolution EUV imagers such as Hi-C \citep{Subramanian2018} and 
the Extreme Ultraviolet Imager (EUI) aboard Solar Orbiter 
\citep{Mueller2020,EUI} have revealed abundant small-scale brightenings in 
the quiet Sun. The discovery of `EUV campfires' by \citet{Berghmans2021} 
demonstrated that impulsive heating events occur ubiquitously in the low 
corona at energies well below those of classical nanoflares. Subsequent 
observations have extended this picture to previously unexplored spatial 
and temporal scales \citep{Chitta2023, Chitta2025} and to less accessible 
low coronal layers \citep{Chitta2021}.

Crucially, this progress does not imply that elementary dissipation 
scales are being resolved. Rather, it constrains the macroscopic plasma 
response -- thermal energy content, characteristic timescales, spatial 
extent, and height distribution -- associated with impulsive heating, 
precisely the regime predicted by nanoflare theory to be observationally 
accessible \citep{Klimchuk2015, Klimchuk2017}. In this work, we analysed EUV campfires from the Solar Orbiter/EUI 
catalogue of  \citet{Berghmans2021}. We performed a systematic analysis of 
their thermodynamic properties, thermal energies, and statistical 
distributions, assessing robustness against geometric assumptions. For context, we compared the campfire population with microflares observed 
by the Spectrometer/Telescope for Imaging X-rays (STIX) on Solar Orbiter 
\citep{Battaglia2021} and with nanoflare-like events observed by the 
Atmospheric Imaging Assembly (AIA) on the Solar Dynamics Observatory (SDO) 
\citep{Purkhart2022}. A key aspect of this study is the layer-resolved 
characterisation of campfire properties across atmospheric layers, 
which provides new constraints on how dissipation timescales vary with plasma 
conditions. We emphasise that our analysis constrains 
observable thermal envelopes rather than the microscopic dissipation 
sites themselves.

\section{Observations and data analysis}

\subsection{Observational setup and dataset}

This study uses the observations obtained on 30~May~2020 by the 
High-Resolution Imager in the extreme ultraviolet (\hrieuv) of the 
EUI aboard Solar Orbiter 
\citep{Mueller2020}. At the time of observation, Solar Orbiter 
was at a heliocentric distance of 0.556~au. With its plate scale of 
$0.49''$~pixel$^{-1}$, compared with $0.6''$~pixel$^{-1}$ for 
the SDO/AIA \citep[][]{Lemen2012}, 
and owing to the reduced solar distance, \hrieuv\ resolves a 
projected pixel scale of $\approx$198~km at the solar surface, about 
2.2 times finer than the $\approx$435~km of AIA at 1~au.

\hrieuv observes in a narrow passband centred at 174~\AA, dominated 
by Fe\,\textsc{ix}/Fe\,\textsc{x} emission and sensitive to plasma 
near 1~MK, characteristic of the quiet-Sun corona. The effective 
pixel scale at the solar surface was 198~km, with a temporal cadence 
of 5~s per frame, sufficient to resolve the rapid evolution of 
transient brightenings.

We analysed EUV campfires from the catalogue of  \citet{Berghmans2021}. 
The image sequence spans 245~s over a field of view of approximately 
$17' \times 17'$ ($\sim1.7\times10^{21}$~cm$^{2}$, or 
$\approx411\times411$~Mm$^{2}$). The observations were obtained near 
solar minimum, with no active regions in the field of view, making 
the dataset well suited for studying weak, small-scale energy-release 
events free from active-region contamination.

All data were calibrated using the standard EUI pipeline described in 
\citet{Berghmans2021} (hereafter Paper~I), including corrections for 
dark current, flat-field, detector degradation, and pointing. No 
additional processing was applied beyond what was presented in Paper~I.

\subsection{Campfire event sample and co-identification}

The campfire sample is drawn from the event catalogue of  
\citet{Berghmans2021}. We used the primary sample of 1,468 campfires 
detected at $\geq5\sigma$.

\begin{figure*}[t!]
\centering
\includegraphics[width=0.8\linewidth]{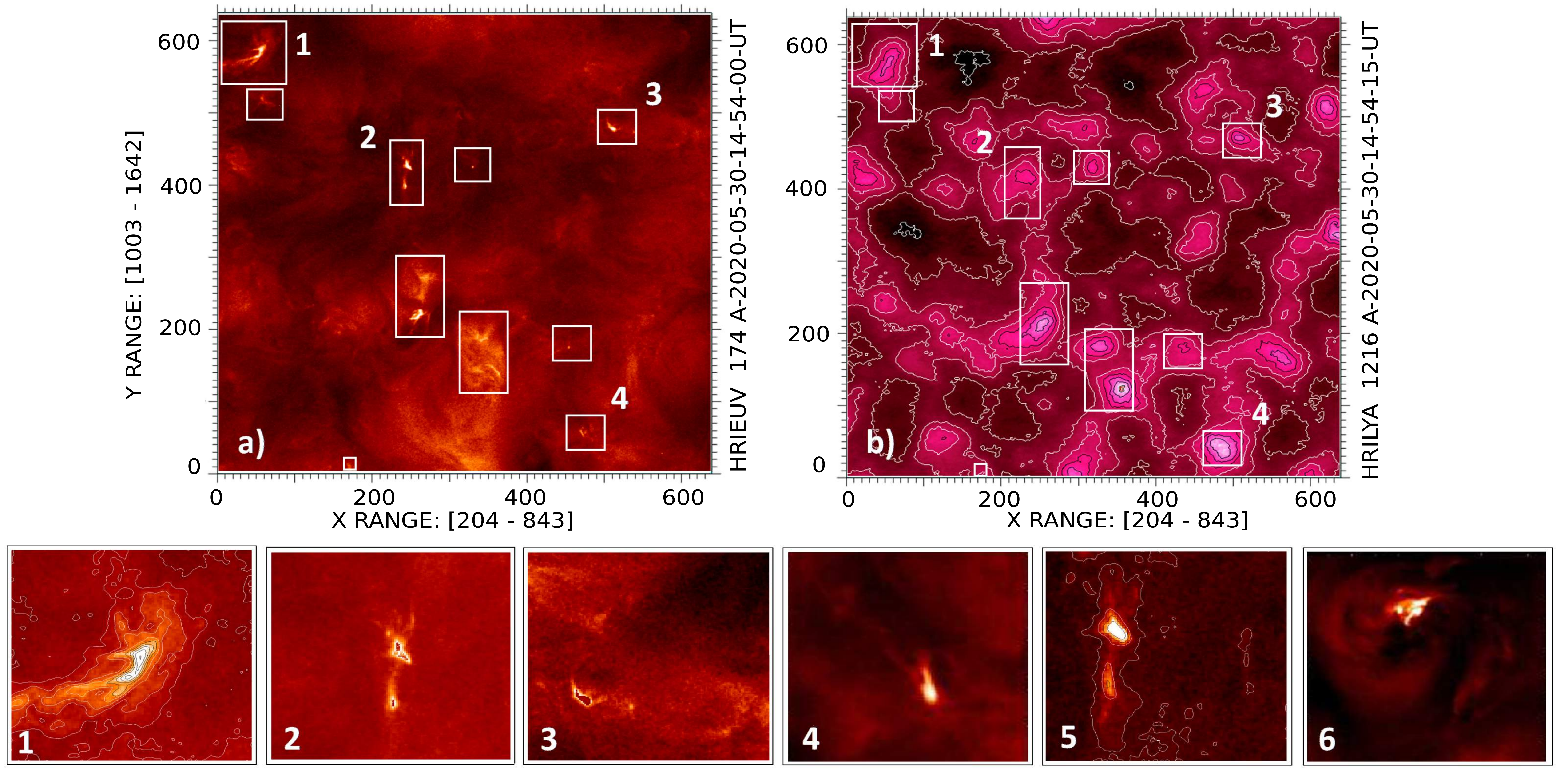}
\caption{Representative EUV campfires from the Solar Orbiter \hrieuv\ 
catalogue of  \citet{Berghmans2021}, observed at 174~\AA\ on 30~May~2020. 
\textit{Panel~(a)}: \hrieuv\ 174~\AA\ field of view with campfire 
locations marked. \textit{Panel~(b)}: Co-temporal \hrilya\ 1216~\AA\ 
image with Ly$\alpha$ network contours. \textit{Bottom row}: Events~1--6. Events 1--4 lie above the 
Ly$\alpha$ network and are marked in panels (a) and (b).  Events~5 
and~6 originate from other quiet-Sun regions. The bottom row illustrates a range of morphologies, 
including compact brightenings, loop-like structures, and events 
with apparent flows. Animations corresponding to panels~(a) and (b) and 
events~1--6 are available online.
}
\label{fig:campfire_examples}
\end{figure*}

All events were co-identified in simultaneous SDO 
\citep{Pesnell2012} AIA 171~\AA\ and 193~\AA\ observations 
\citep{Lemen2012}, from which thermodynamic parameters (temperature 
and emission measure) were derived in Paper~I using differential 
emission measure (DEM) analysis. The same co-identification enabled 
subsequent stereoscopic height determinations in Paper~II 
\citep{Zhukov2021}. Throughout this paper, `EUV campfires' refers 
exclusively to the events catalogued by \citet{Berghmans2021}. 
Figure~\ref{fig:campfire_examples} shows representative campfire 
events, illustrating their morphology, segmentation, and the geometric 
parameters used in this study.

\subsection{Event definition and geometric characterisation}

Event identification, tracking, and lifetime determination follow 
the definitions of Paper~I. An event is a spatially connected set 
of pixels persisting over consecutive frames, with its lifetime 
measured between the first and last frames exceeding the detection 
threshold; the minimum detectable duration corresponds to a single 
5~s frame. Energetics are characterised using the peak values of 
emission measure and temperature reached over the event lifetime 
rather than integrated over the full duration. At this stage, the 
spatial extent is quantified using geometrical descriptors derived 
from the intensity distribution:
the projected area, the major axis ($L$) and minor axis ($W$) from second moments,
and the bounding-box dimensions.

\section{Thermal energy framework}

\subsection{Thermal energy formulation}

We focused exclusively on the thermal component of the energy budget 
inferred from EUV observations, combining thermodynamic diagnostics 
with geometrical source characterisation and explicit background 
subtraction. For an optically thin coronal plasma, the thermal energy 
of a heating event is
\[
E_{\mathrm{th}} = 3\,k_{\mathrm{B}}\,T\,n_{\mathrm{e}}\,V,
\]
where $k_{\mathrm{B}}$ is the Boltzmann constant, $T$ the plasma 
temperature, $n_{\mathrm{e}}$ the electron number density, and $V$ 
the emitting volume. For a fully ionised hydrogen plasma, 
$n_{\mathrm{tot}} \approx 2\,n_{\mathrm{e}}$, and the factor 3 
accounts for both electrons and ions \citep{Aschwanden2000}.

Since $n_{\mathrm{e}}$ is not directly observable, we express the 
thermal energy in terms of the emission measure. DEM analysis yields 
the column emission measure,
\[
\mathrm{EM}_{\mathrm{col}} = \int n_{\mathrm{e}}^{2}\, \mathrm{d}s,
\]
integrated along the line of sight. Given an effective emitting depth 
$h_{\mathrm{eff}}$ and projected source area $A$, the volume emission 
measure is
\[
\mathrm{EM}_{\mathrm{vol}} = \mathrm{EM}_{\mathrm{col}}\,A,
\]
and the thermal energy becomes
\[
E_{\mathrm{th}} = 3\,k_{\mathrm{B}}\,T\,A\,
\sqrt{\mathrm{EM}_{\mathrm{col}}\,h_{\mathrm{eff}}}
= 3\,k_{\mathrm{B}}\,T\,\sqrt{\mathrm{EM}_{\mathrm{vol}}\,V},
\]
which follows from $\mathrm{EM}_{\mathrm{vol}} = n_{\mathrm{e}}^{2}\,V$ 
and $E_{\mathrm{th}} = 3\,n_{\mathrm{e}}\,k_{\mathrm{B}}\,T\,V$. 
For the reference energy estimates, we adopted
$EM_{\mathrm{vol}} = EM_{\mathrm{col}} A_{\mathrm{proj}}$
with the baseline effective volume $V_{\mathrm{proj}}$
defined in Sect.~\ref{Geometry}.

In practice, $E_{\mathrm{th}}$ was computed from background-subtracted 
peak values of emission measure and temperature -- $\mathit{EM}_{\max}$ 
and $T_{\max}$ -- evaluated at the single time step of maximum 
spatially integrated EUV intensity. This instantaneous estimate is 
combined with the emitting volume at the same time step, and is fully 
equivalent to using the volume emission measure directly. It enables 
comparisons with previous EUV and X-ray studies 
\citep{Aschwanden2000, Ulyanov2019} and forms the basis for all 
subsequent energy estimates. As the plasma undergoes ongoing radiative 
and conductive losses throughout the event lifetime, this peak thermal 
energy represents a conservative estimate of the total heating requirement; 
radiative losses are addressed quantitatively in 
Sect.~\ref{sec:heating_vs_radiation}. Geometric volume assumptions are 
explored in Sect.~\ref{Geometry}.

\subsection{Thermodynamic parameters and background subtraction}

Temperatures and column emission measures were adopted from Paper~I, 
where they were derived via DEM 
analysis using the regularised inversion code of 
\citet{Hannah2012} applied to co-temporal SDO/AIA 
multi-channel observations and 
validated through cross-channel consistency checks. The Spectral Imaging of 
the Coronal Environment (SPICE) instrument aboard Solar Orbiter was not yet 
operational on 30~May~2020, making 
SDO/AIA the only available multi-temperature diagnostic. 
Re-analyses using AIA time-lag techniques \citep{Dolliou2023} and
SPICE observations of similar events \citep{Dolliou2024, Huang2023}
are discussed in Sect.~\ref{sec:comparison}.

In the quiet Sun, a substantial fraction of the observed EUV emission 
originates from pre-existing coronal plasma unrelated to the impulsive 
event. To isolate the transient thermal signature, we computed both raw 
and background-subtracted thermodynamic parameters. For each 
campfire, the DEM was obtained using the regularised inversion 
method of \citet{Hannah2012} applied to the $5\sigma$ detections. 
For the background-subtracted data, the background immediately 
preceding each event was estimated using the inpainting method of 
\citet{Bertalmio2001}, which estimates the intensity and intensity 
gradient within the masked region to match the surrounding 
boundary, and removed prior to the DEM inversion.
Background-subtracted values represent the excess thermal plasma 
produced by the impulsive episode; non-subtracted values reflect the 
total emitting structure along the line of sight. Unless stated 
otherwise, background-subtracted parameters were used for all energy 
estimates, consistent with standard practice in flare and nanoflare 
studies. Of the 1,468 campfires, 558 have valid DEM solutions 
without background subtraction, and 389 retain robust solutions 
after background subtraction; only these 389 events 
($\sim$26\% of the sample) were included in the energy analysis.

Throughout this work, the height of each event refers to 
its stereoscopic altitude above the solar surface, determined in 
Paper~II \citep{Zhukov2021} from the simultaneous SDO and 
Solar Orbiter vantage points. Events were grouped into atmospheric 
layers according to this stereoscopic height, following 
\citet{Judge2006} (see our Table~\ref{tab:campfire_layer_summary}). This 
geometric height gives the vertical position of the event in the 
atmosphere and is distinct from the effective emitting depth ($h_{\mathrm{eff}}$) along the line of sight, introduced earlier in 
the thermal energy formulation. The effective depth $h_{\mathrm{eff}}$ and 
volume $V$ cannot be measured directly and are constrained via the 
geometrical models introduced in Sect.~\ref{Geometry}.

The fraction of events retaining a robust DEM solution shows no 
systematic trend with height (ranging from $\sim$22\% to $\sim$39\% 
across layers, without a monotonic dependence), confirming that the 
DEM-based subsample is not preferentially biased towards any 
particular atmospheric layer. The height-resolved comparison of 
thermal properties is therefore unaffected by DEM completeness. 
The subsample is, however, biased towards larger and longer-lived 
events, for which the emission-measure signal is sufficient for a 
stable inversion; energy-based conclusions accordingly refer to 
this brighter population. We stress that the duration--height 
relation itself is derived from the full sample of 1,468 events, 
for which durations and heights are available independently of the 
DEM inversion, and is therefore not affected by this selection. 
The thermal-energy comparison across layers, by contrast, uses the 
389-event subsample; within this subsample the energies show at 
most a weak dependence on height, much smaller than the variation 
in duration measured on the full sample.

\subsection{Thermodynamic properties and \texorpdfstring{$\mathit{EM}$--$T$}{EM--T} scaling}

\begin{figure}[t!]
\centering
\includegraphics[width=0.85\linewidth]{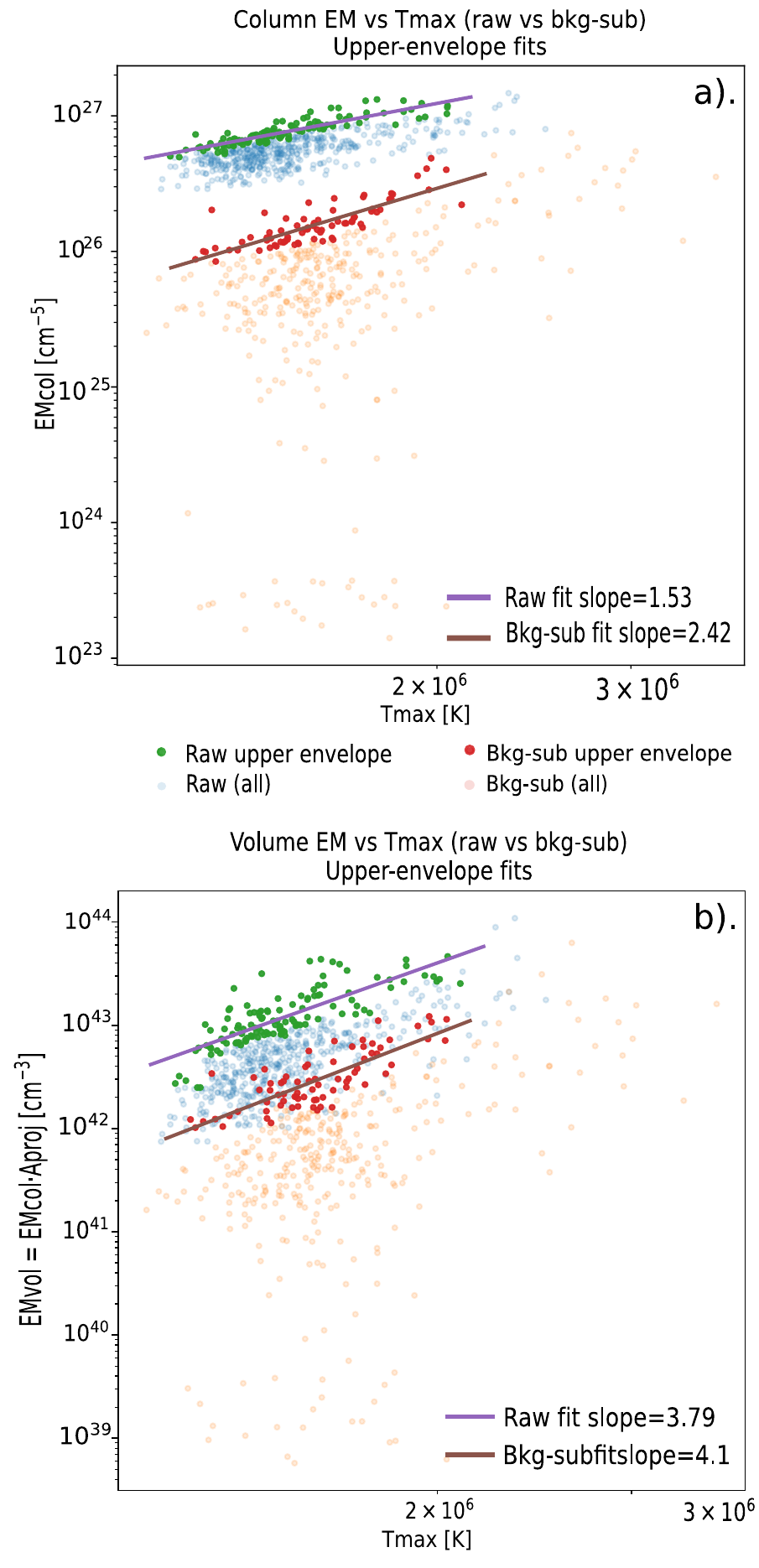}
\caption{Emission measure--temperature ($\mathit{EM}$--$T$) relations for 
campfires.
\textit{Top}: Column emission measure ($\mathit{EM}_{\mathrm{col}}$)  versus temperature. \textit{Bottom}: Volume emission measure, 
$\mathit{EM}_{\mathrm{vol}} = \mathit{EM}_{\mathrm{col}}\,A_{\mathrm{proj}}$. 
Blue and orange points show all events without and with background 
subtraction, respectively. Green and red filled circles mark the 
upper-envelope points; purple and brown lines show power-law fits 
to the raw and background-subtracted upper envelopes, with slopes 
indicated. The upper envelope -- the locus of maximum $\mathit{EM}$ 
at each temperature -- is fitted rather than the full distribution 
because it traces the most energetic events, is less sensitive to 
detection incompleteness at low energies, and enables direct 
comparisons with published $\mathit{EM}$--$T$ scaling relations 
\citep[e.g.][]{Aschwanden2004, Warmuth2016A&A}. 
The two fits shown in each panel correspond to the raw and 
background-subtracted upper envelopes: for $\mathit{EM}_{\mathrm{col}}$ 
the slopes are $1.53$ (raw) and $2.42$ (background-subtracted), 
and for $\mathit{EM}_{\mathrm{vol}}$ they are $3.79$ (raw) and 
$4.1$ (background-subtracted).
}
\label{fig:emt_scaling}
\end{figure}

The emission measure--temperature ($\mathit{EM}$--$T$) plane 
provides a compact characterisation of the thermal plasma response 
to impulsive energy release. Temperature describes how strongly the 
emitting plasma is heated, whereas the emission measure quantifies 
the amount of plasma contributing to the observed emission. In X-ray 
flare studies these quantities are routinely derived from spectral 
or broadband measurements and used to compare the thermal properties 
of events over a wide range of magnitudes 
\citep[e.g.][]{Aschwanden2004, Warmuth2016A&A}. Here, the 
DEM analysis of the co-temporal SDO/AIA 
observations allows EUV campfires to be placed in the same 
thermodynamic parameter space, testing whether they extend the 
empirical $\mathit{EM}$--$T$ scaling of nanoflares and larger flares 
towards lower temperatures and emission measures 
\citep[e.g.][]{Joulin2016, Purkhart2022}. The comparison concerns 
the thermal plasma response only and does not constrain a 
non-thermal particle component.

In Fig.~\ref{fig:emt_scaling}, each point corresponds to a 
single campfire. The values used are the maximum column emission 
measure and the maximum DEM-weighted temperature 
\citep[following the definition of][]{Parenti2017}, taken over all 
spatial pixels and time steps belonging to each campfire and 
determined independently, so that the two maxima need not occur at 
the same pixel or time; both are derived from the DEM maps of 
Paper~I \citep{Berghmans2021}, which employs the DEM method of 
\citet{Hannah2012}. The volume emission measure follows as 
$\mathit{EM}_{\mathrm{vol}} = \mathit{EM}_{\mathrm{col}}\,A_{\mathrm{proj}}$. 
Values obtained with and without background subtraction are shown 
separately, and the background-subtracted values are used for all 
energy estimates. The upper-envelope points (green and red) trace 
the maximum $\mathit{EM}$ reached as a function of temperature, and 
it is these that the power-law fits describe.

Campfires show temperatures of $T \sim 1$--$2$~MK, consistent with
their detection in the EUI~174~\AA\ passband. Their emission measures 
span several orders of magnitude, reflecting a broad range of event 
energetics. Recent analyses suggest that a significant fraction of 
campfire plasma does not reach coronal temperatures 
\citep{Dolliou2023, Dolliou2024, Huang2023}; the thermal energies 
derived here therefore characterise only the coronal ($\sim$1~MK) 
component visible in this passband.

Power-law fits to this upper envelope are overplotted in 
Fig.~\ref{fig:emt_scaling}. For the background-subtracted data, 
the upper-envelope slopes are $2.42$ for $\mathit{EM}_{\mathrm{col}}$ 
and $4.1$ for $\mathit{EM}_{\mathrm{vol}}$. These values are 
comparable to those reported for impulsively heated coronal loops, 
nanoflares, and larger flares 
\citep[e.g.][]{Aschwanden2004, Warmuth2016A&A}; in particular, 
the $\mathit{EM}_{\mathrm{vol}} \propto T^{4.1}$ scaling is 
consistent with the $\mathit{EM} \propto T^{4}$ relation found for 
larger flares \citep{Warmuth2016A&A}, supporting the interpretation 
of campfires as the low-energy extension of impulsive heating in 
magnetically confined plasma.

\subsection{Geometrical volume models}
\label{Geometry}

We considered four effective volume models spanning the range commonly 
used in nanoflare and EUV energetics studies, all evaluated at the 
time of peak emission. Models $V_{\mathrm{proj}}$, $V_{\mathrm{cyl}}$, 
and $V_{\mathrm{box}}$ include a filling factor of 0.1, consistent 
with values adopted in previous studies \citep{Aschwanden2000, 
Ulyanov2019} and supported by an independent check: comparing 
$V_{\mathrm{EUV}}$ with $V_{\mathrm{box}}$ yields filling-factor 
ratios of 0.07--0.17 (mean $\approx$0.13). The $V_{\mathrm{EUV}}$ 
model intrinsically represents the EUV-emitting fraction and requires 
no additional filling factor.

The area--depth scaling ($V_{\mathrm{proj}}$; a scale-free proxy)
is\[
V_{\mathrm{proj}} = 0.1 \, A_{\mathrm{proj}} \, \sqrt{A_{\mathrm{proj}}},
\]
where $A_{\mathrm{proj}}$ is the projected emitting area at peak 
emission. The line-of-sight depth scales with $\sqrt{A_{\mathrm{proj}}}$, 
yielding a scale-free estimate applicable when no specific loop 
geometry is assumed. This is the baseline model used throughout unless 
stated otherwise.

The cylindrical loop volume ($V_{\mathrm{cyl}}$) 
is\[
V_{\mathrm{cyl}}= 0.1 \, \pi \left( \frac{W}{2} \right)^2 L,
\]
where $L$ and $W$ are the major and minor axes from second-moment 
analysis. This model assumes a straight cylindrical loop with 
circular cross-section.

The bounding-box scaling ($V_{\mathrm{box}}$) is\[
V_{\mathrm{box}} = 0.1 \, A_{\mathrm{box}} \, \sqrt{A_{\mathrm{box}}},
\]
where $A_{\mathrm{box}} = L_{\mathrm{box}} W_{\mathrm{box}}$ is the 
minimal bounding-rectangle area. This model accommodates irregular 
or multi-stranded morphologies.

The Aschwanden-type elliptical-loop EUV volume ($V_{\mathrm{EUV}}$) is
\[
V_{\mathrm{EUV}}= V_{\mathrm{loop}}
\left[
    1 - \frac{2}{\pi}
    \arctan\!\left(\frac{h_{\mathrm{ch}}}{L/2}\right)
\right],
\qquad
V_{\mathrm{loop}} = \frac{\pi^{2}}{4}\, r\, w^{2},
\]
with
\[
r = \sqrt{\left(\frac{L}{2}\right)^2 + h_{\mathrm{ch}}^2} - \frac{w}{2},
\]
where $L$ and $w$ are the ellipse major and minor axes, and 
$h_{\mathrm{ch}}$ is the chromospheric height below which EUV 
emission is suppressed. Classical nanoflare studies adopt 
$h_{\mathrm{ch}} \sim 2500$~km \citep[e.g.][]{Aschwanden2000}; 
here we used $h_{\mathrm{ch}} = 150$~km, motivated by stereoscopic 
measurements that show that campfires form at exceptionally low 
altitudes, often well below 1~Mm \citep{Zhukov2021}. We note that 
$h_{\mathrm{ch}}$ is a physical opacity cutoff, not a minimum 
campfire height; the observed height distribution 
(Fig.~\ref{fig:composite_judge2006}) shows events typically above 
$\sim$500~km, while $h_{\mathrm{ch}}$ describes the layer below 
which EUV emission is suppressed regardless of event location. This 
model describes the EUV-bright coronal portion of a semi-circular 
loop \citep{Aschwanden2000} and requires no filling factor.

\subsection{Scope and limitations}
The thermal energies derived here represent only the excess thermal 
component inferred from EUV emission, and the statistical properties 
of the campfire population -- energy range, distribution shape, and 
scaling behaviour -- are robust against reasonable variations in 
geometry and background subtraction. A quantitative assessment of 
geometric uncertainties is given in Sect.~\ref{sec:uncertainty}, and 
the broader limitations of the analysis are discussed in 
Sect.~\ref{sec:limitations}.

\section{Results}

We first characterised the morphology and source geometry of the 
campfire population, then quantified their thermal energies and 
statistical distributions, and finally examined how these properties 
vary with atmospheric height. Throughout, statistical distributions 
are characterised using complementary cumulative distribution 
functions (CCDFs) and power-law slopes estimated via maximum 
likelihood, which together provide robust, binning-independent 
diagnostics of the underlying populations.

\subsection{Morphology and source geometry}

\begin{figure*}[t!]
\sidecaption
\includegraphics[width=0.67\textwidth]{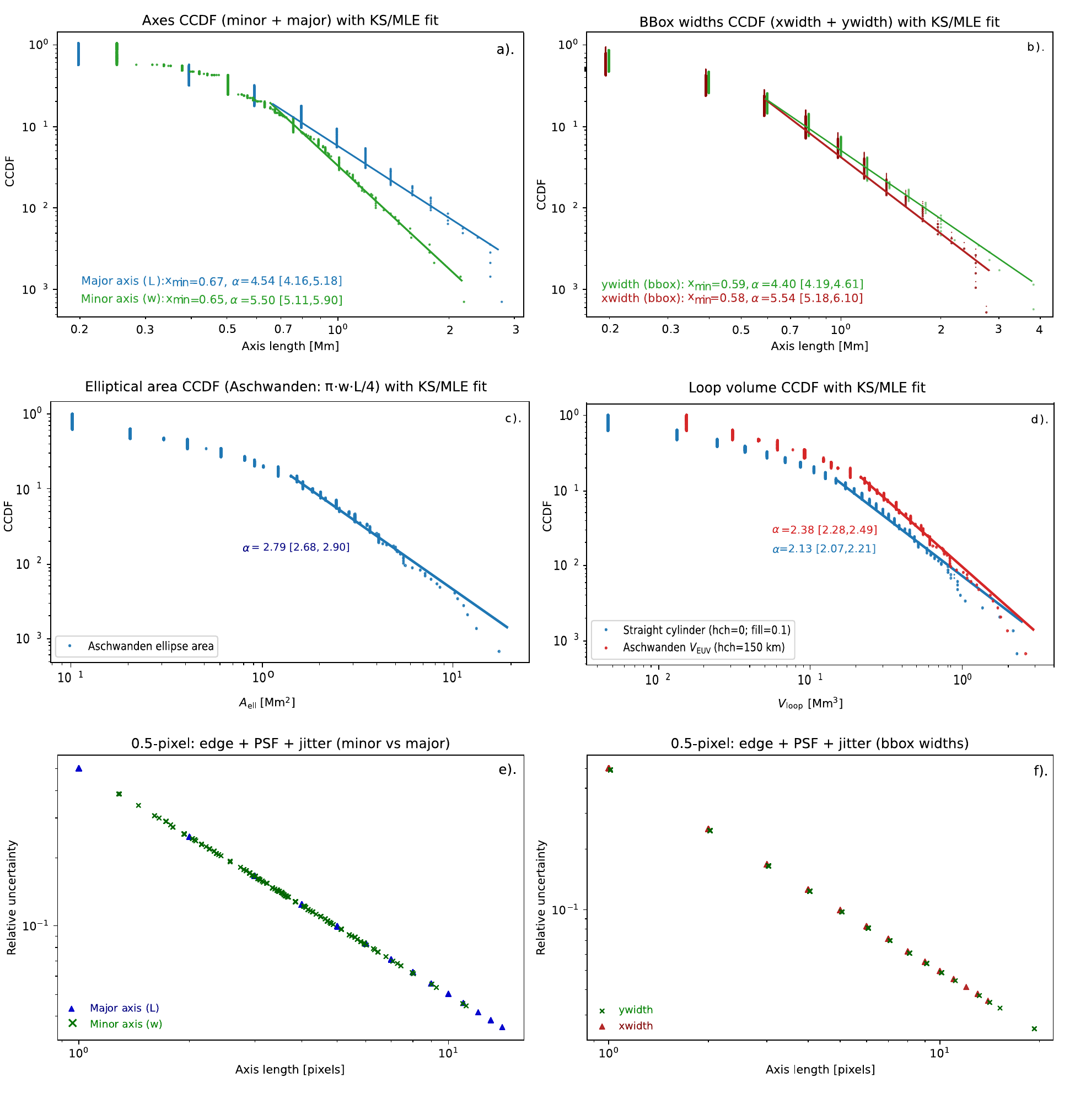}
\caption{CCDFs of campfire morphological and geometric parameters at peak 
emission. \textit{Panel (a)}:~Minor axis ($W$; blue) and major axis ($L$; green).  \textit{Panel (b)}:~Bounding-box widths in the $x$- (red) and $y$-direction (green).  \textit{Panel
(c)}:~Elliptical projected area $A_{\mathrm{ell}} = \pi\,W\,L/4$. 
\textit{Panel (d)}:~Effective loop volume for a straight-cylinder model (blue) and 
the Aschwanden $V_{\mathrm{EUV}}$ model (red). 
\textit{Panels (e) and (f)}: Relative uncertainties as a function of axis length and 
bounding-box width in pixels under a conservative half-pixel error; 
in each panel the two parameters are shown with different symbols 
(triangles and crosses) for clarity. 
The concentration near $\sim$0.2~Mm in (a) and (b) reflects the 
minimum detectable size of one pixel ($\approx$198~km). Power-law 
fits to the high-value tails are shown in (a)--(d), with the lower 
cutoff ($x_{\min}$) from the KS criterion and
indices $\alpha$ via maximum likelihood. Vertical bars show 
bootstrap confidence intervals (16th--84th percentiles).
}
\label{fig:morph_ccdf}
\end{figure*}

\begin{figure*}[t!]
\sidecaption
\includegraphics[width=0.67\textwidth]{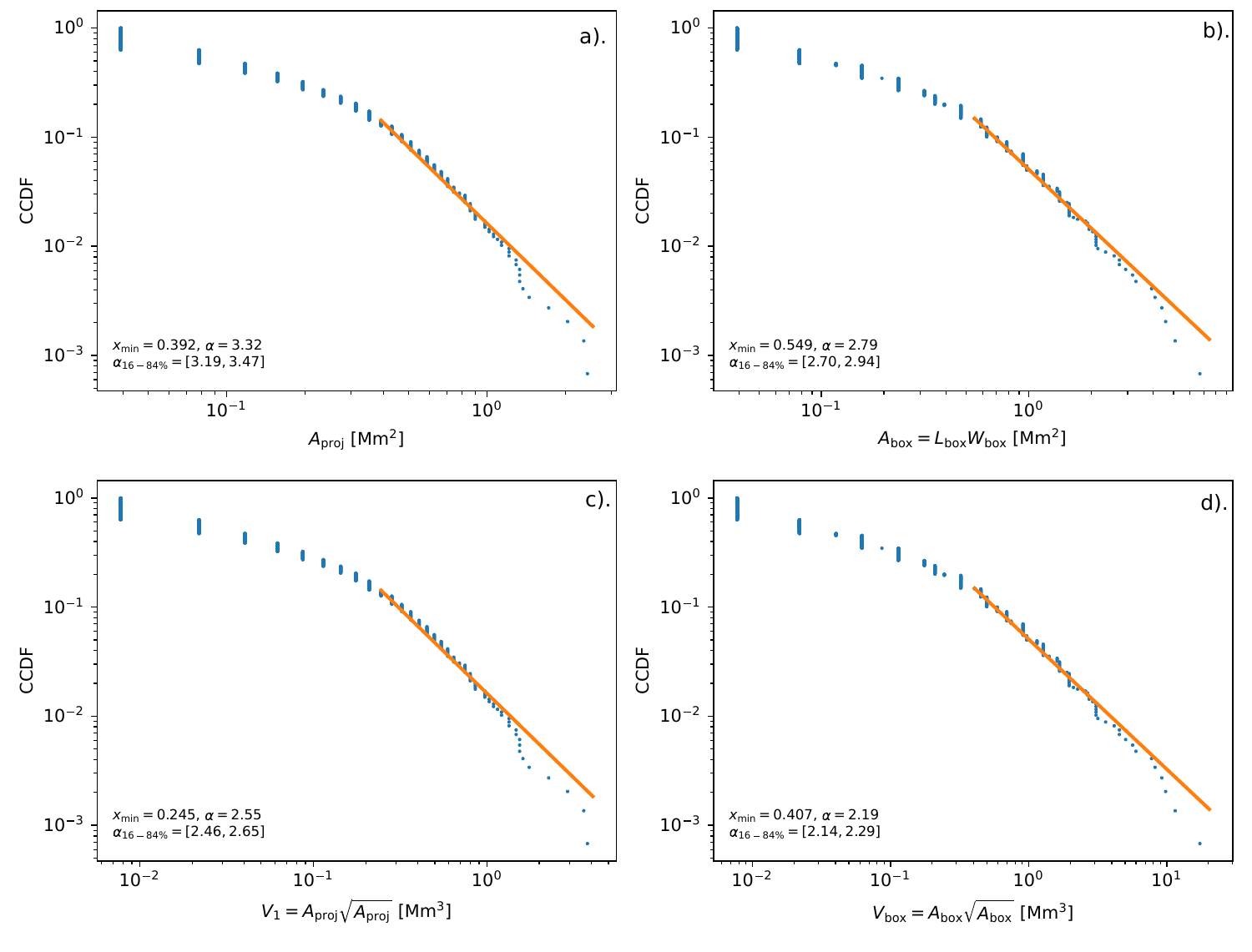}
\caption{CCDFs of campfire geometric quantities: (a)~projected area ($A_{\mathrm{proj}}$), (b)~bounding-box area (
$A_{\mathrm{box}} = L_{\mathrm{box}} W_{\mathrm{box}}$), 
(c)~effective volume ($V_{1} = A_{\mathrm{proj}}\sqrt{A_{\mathrm{proj}}}$), 
and (d)~$V_{\mathrm{box}} = A_{\mathrm{box}}\sqrt{A_{\mathrm{box}}}$. 
Power-law fits are shown for the high-value tails above the 
KS-determined cutoff, $x_{\min}$. Slopes ($\alpha$) are estimated 
via maximum likelihood with bootstrap confidence intervals 
($\alpha_{16-84\%}$); the values are indicated in each panel.
}
\label{fig:ccdf_geom_4panel}
\end{figure*}

Figures~\ref{fig:morph_ccdf} and \ref{fig:ccdf_geom_4panel} show 
CCDFs of campfire morphological parameters and geometric volumes 
evaluated at peak emission, where the CCDF $P(X \geq x)$ gives the 
fraction of events with a value greater than or equal to $x$.
Projected source areas are computed 
assuming an elliptical geometry defined by second-moment axes, 
following an Aschwanden-type formulation that avoids assumptions 
of cylindrical symmetry.

All geometric quantities exhibit power-law tails towards larger 
values, though the fitted regime spans less than one decade above 
$x_{\min}$. Slopes vary across geometric models but remain 
consistent within their bootstrap uncertainties -- confidence 
intervals obtained by repeatedly resampling the data with 
replacement and refitting, which quantify how sensitive the slope 
is to the particular set of detected events. The slopes are also 
stable against conservative assumptions about spatial resolution, 
segmentation, and pointing jitter.

Power-law indices are estimated via maximum-likelihood estimation (\citealt{Clauset2009, verbeeck2019solar}), with the lower cutoff 
$x_{\min}$ selected objectively via the Kolmogorov--Smirnov (KS) 
criterion \citep{Clauset2009}. This approach yields unbiased slope 
estimates and enables direct comparisons with nanoflare and flare 
studies, where power-law slopes are a key diagnostic of coronal 
heating \citep{Hudson1991}. We do not claim that a power law is the 
unique description of the data. Following \citet{Clauset2009}, we first verified the power-law 
hypothesis itself using a bootstrap KS 
goodness-of-fit test ($10^{3}$ synthetic samples): the fit is not 
rejected ($D = 0.05$, $p = 0.60$), so a power law is a plausible 
description of the tail. We then compared it against 
alternative distributions using normalised likelihood ratios. A 
pure exponential is rejected ($R = 2.6$, $p = 0.008$), confirming 
that the distribution is genuinely heavy-tailed. A log-normal, by 
contrast, cannot be statistically distinguished from a power law 
over the limited dynamic range probed here ($p = 0.33$), consistent 
with the observational finding of \citet{verbeeck2019solar} that 
solar-flare energy distributions can be equally well described by a 
log-normal. Such behaviour is expected in self-organised critical 
systems \citep{Bak1987SOC}, where avalanche-like energy release 
follows a power law when events are small compared with the system 
size but departs towards a log-normal as the largest events approach 
the system scale \citep{Vlahos1995, Georgoulis1996}. Crucially, the 
physically relevant quantity for coronal-heating arguments -- the 
slope of the high-value tail ($\alpha \approx 2.2$) -- is robust to 
the choice between these forms; our conclusions rest on the 
heavy-tailed character and tail slope of the distribution rather 
than on a specific functional model.

\subsection{Uncertainty propagation}
\label{sec:uncertainty}

Geometric uncertainties arise primarily from finite spatial 
resolution and segmentation. We adopted a conservative total 
uncertainty of $\sigma_{\mathrm{pix}} = 0.5$~pixel ($\approx$100~km) 
per length scale, combining contributions from edge definition, 
point-spread function, and pointing jitter. Although the 
Nyquist-limited resolution of \hrieuv\ is 2~pixels ($\approx$400~km), 
moment-based axes $L$ and $W$ are computed from intensity-weighted 
second moments over the full event area, reducing the effective 
uncertainty well below one pixel for all but the smallest events. 
This yields relative uncertainties of $\sim$1--3\% for large events 
and $\sim$3--7\% for typical events (see the bottom row of 
Fig.~\ref{fig:morph_ccdf}).

For volume models of the form $V \propto W^{2} L$,
\[
\frac{\sigma_{V}}{V} =
\sqrt{\left(2\frac{\sigma_{W}}{W}\right)^{2}
      + \left(\frac{\sigma_{L}}{L}\right)^{2}}.
\]
Since $E_{\mathrm{th}} \propto \sqrt{V}$ (from 
$E_{\mathrm{th}} = 3\,k_{\mathrm{B}}\,T\,n_e\,V$ and 
$n_e \propto \sqrt{\mathrm{EM}_{\mathrm{vol}}/V}$), thermal-energy 
uncertainties remain modest relative to the several 
orders-of-magnitude dynamic range of the observed distributions.

\subsection{Thermal energy distributions}

\begin{figure}[t!]
\centering
\includegraphics[width=1.0\linewidth]{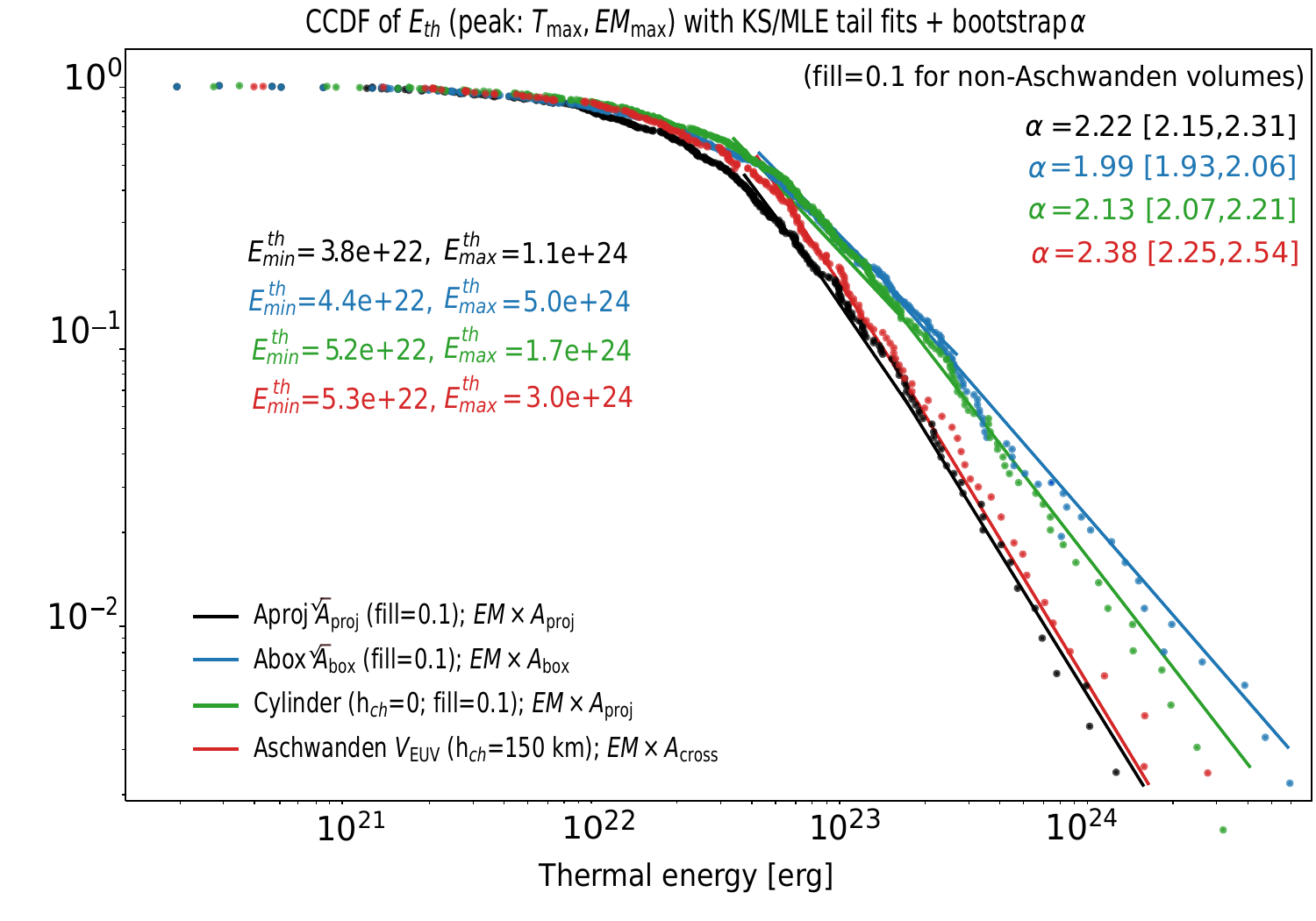}
\caption{CCDFs of campfire thermal energy ($E_{\mathrm{th}}$) for four geometric 
volume models: projected area ($A_{\mathrm{proj}}$; black), 
bounding-box area ($A_{\mathrm{box}}$; blue), cylindrical geometry 
(green), and Aschwanden-type EUV loop volume (red). A filling factor 
of 0.1 is applied for all non-Aschwanden volumes. Power-law fits are 
overplotted on the data above the KS-determined cutoff, $E_{\min}$; 
bootstrap 16th--84th percentile confidence intervals on $\alpha$ are 
quoted. The fitted energy range $[E_{\min}, E_{\max}]$ 
is indicated for each model. Slopes $\alpha \approx 2.0$--$2.4$ are 
consistent across all models, demonstrating robustness to geometric 
assumptions.
}
\label{fig:eth_ccdf_models}
\end{figure}

\begin{figure}[t!]
\centering
\includegraphics[width=1.0\linewidth]{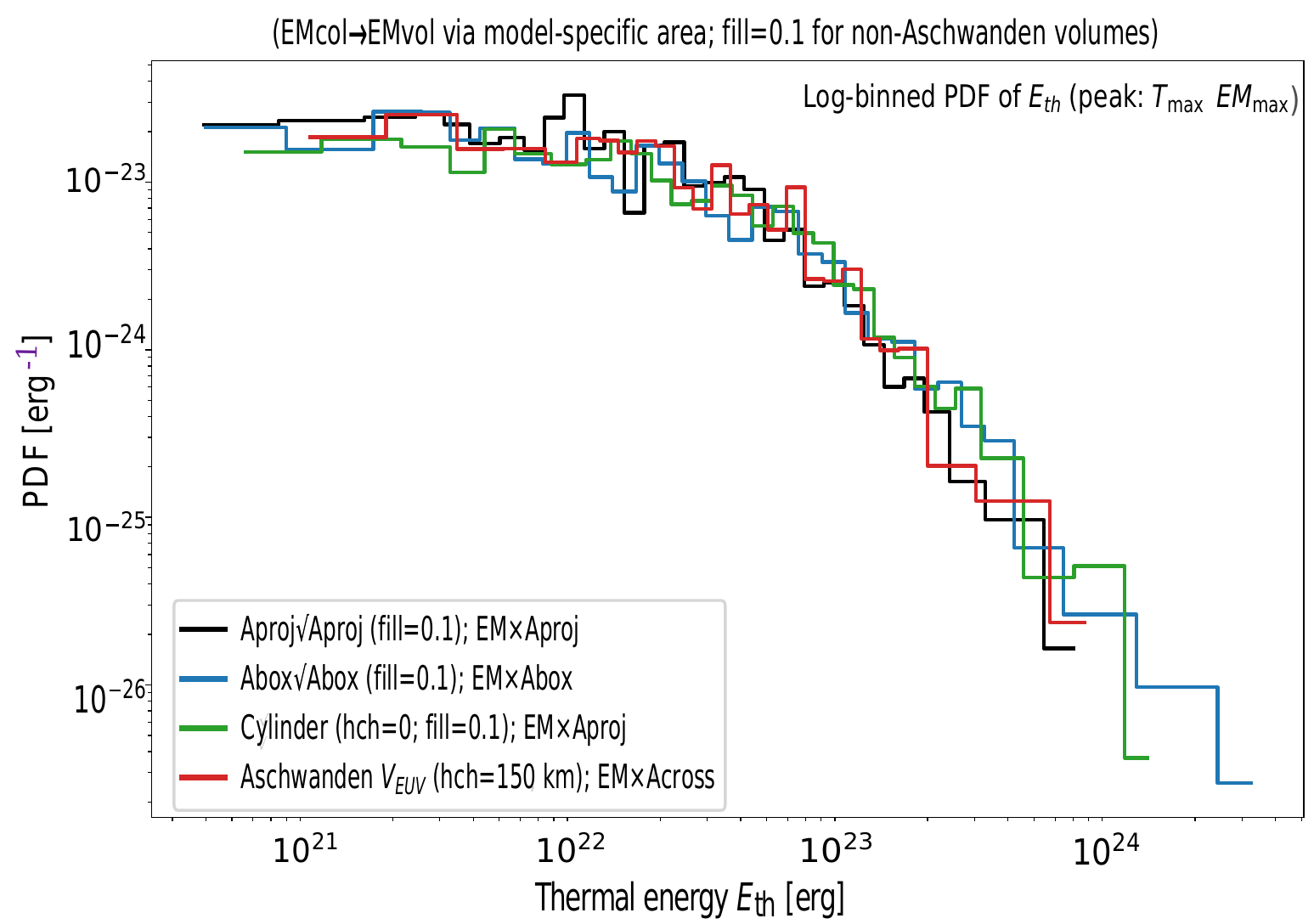}
\caption{Log-binned PDFs of campfire thermal energies for four volume models: 
$A_{\mathrm{proj}}$ (black), $A_{\mathrm{box}}$ (blue), cylindrical 
(green), and Aschwanden-type EUV loop (red). A filling factor of 0.1 
is applied for all non-Aschwanden volumes. PDFs are shown for visual 
comparison; quantitative slope estimates are derived from the CCDFs 
(Fig.~\ref{fig:eth_ccdf_models}).
}
\label{fig:eth_pdf_models}
\end{figure}

The slope of the high-energy tail of the thermal-energy distribution 
is a key diagnostic for coronal heating: it determines whether the 
cumulative energy budget is dominated by rare large events or by 
numerous small ones \citep{Hudson1991}.  As above, we used CCDFs 
throughout for their binning-independent robustness 
\citep{Clauset2009}.

Figure~\ref{fig:eth_ccdf_models} shows thermal-energy CCDFs for 
the four volume models of Sect.~\ref{Geometry}. Slopes were estimated by maximum likelihood above an objectively determined 
$E_{\min}$, with bootstrap confidence intervals obtained from 1000 resamplings. The inferred slopes 
cluster around $\alpha \approx 2.2$ for $A_{\mathrm{proj}}^{3/2}$-scaled 
models, with the Aschwanden model yielding a slightly steeper value 
reflecting its different geometry and the reduced chromospheric 
cutoff $h_{\mathrm{ch}} = 150$~km adopted here. The slopes span a range of 
$\alpha \approx 2.0$--$2.4$, reflecting differences in the adopted geometry; 
the scale-free character of the energy distribution is robust to geometric assumptions.

Complementary log-binned probability density functions (PDFs) are shown in Fig.~\ref{fig:eth_pdf_models}
for comparison with earlier frequency-distribution studies. They were 
not used for slope estimation due to sensitivity to binning choices.

Campfire thermal energies extend to $\sim5\times10^{24}$~erg and 
follow a power law with $\alpha > 2$. Events near this upper end 
are intrinsically rare, as reflected in the CCDF tails. This places 
campfires in the regime where small-scale events can in principle 
contribute significantly to coronal heating, if the power law 
extends below current detection limits.

\begin{table*}[t]
\centering
\caption{Campfire properties by atmospheric layer.}
\label{tab:campfire_layer_summary}
\begin{tabular}{lccccccccc}
\hline
Layer & Height range & $N$ & Peak height & $t_{\min}$ & $t_{90}$ & $t_{95}$ & 
$V_{\max}$ & $E_{\rm th,16}$ & $E_{\rm th,84}$ \\
 & (Mm) & & (Mm) & (s) & (s) & (s) & (cm$^{3}$) & (erg) & (erg) \\
\hline
Photosphere & 0.0--0.6 & 14 & 0.4 & $\sim$10 & $\sim$19 & $\sim$20 & $\sim9\times10^{21}$ & $\sim9\times10^{20}$ & $\sim7\times10^{21}$ \\
Lower chromosphere & 0.6--1.2 & 61 & 1.0 & $\sim$10 & $\sim$45 & $\sim$80 & $\sim4\times10^{23}$ & $\sim2\times10^{22}$ & $\sim10^{23}$ \\
Upper chromosphere & 1.2--1.5 & 62 & 1.4 & $\sim$10 & $\sim$39 & $\sim$45 & $\sim3\times10^{23}$ & $\sim2\times10^{22}$ & $\sim2\times10^{23}$ \\
TR + low corona & 1.5--2.0 & 214 & 1.8 & $\sim$10 & $\sim$35 & $\sim$40 & $\sim2\times10^{23}$ & $\sim10^{22}$ & $\sim2\times10^{23}$ \\
Corona ($\beta < 1$) & 2.0--2.4 & 217 & 2.2 & $\sim$10 & $\sim$35 & $\sim$55 & $\sim10^{23}$ & $\sim2\times10^{22}$ & $\sim2\times10^{23}$ \\
Corona ($\beta \ll 1$) & $>$2.4 & 900 & 3.2 & $\sim$10 & $\sim$25 & $\sim$40 & $\sim4\times10^{23}$ & $\sim8\times10^{21}$ & $\sim10^{23}$ \\
\hline
\end{tabular}
\tablefoot{Atmospheric layers follow \citet{Judge2006}. Columns are: layer name; height range; 
number of events $N$; median peak height; minimum duration 
$t_{\min}$; the 90th and 95th percentiles of the duration 
distribution, $t_{90}$ and $t_{95}$; the maximum emitting volume 
$V_{\max}$; and the 16th and 84th percentiles of the thermal energy, 
$E_{\rm th,16}$ and $E_{\rm th,84}$. Durations assume the 5~s 
\hrieuv\ cadence; the shortest events may be single-frame detections 
($t \lesssim 5$~s), with a two-frame minimum of 
$t_{\min} \approx 10$~s. The percentiles $t_{90}$ and $t_{95}$ 
characterise the typical upper duration while remaining insensitive 
to rare long-lived outliers; the full distribution tails are shown 
in Fig.~\ref{fig:composite_judge2006}b. Volumes and energies use 
background-subtracted DEM values with 
$EM_{\rm vol} = EM_{\rm col} A_{\rm proj}$ and 
$V_{\rm proj} = 0.1\,A_{\rm proj}\sqrt{A_{\rm proj}}$.}

\end{table*}

\begin{figure*}[t!]
\centering
\includegraphics[width=\linewidth]{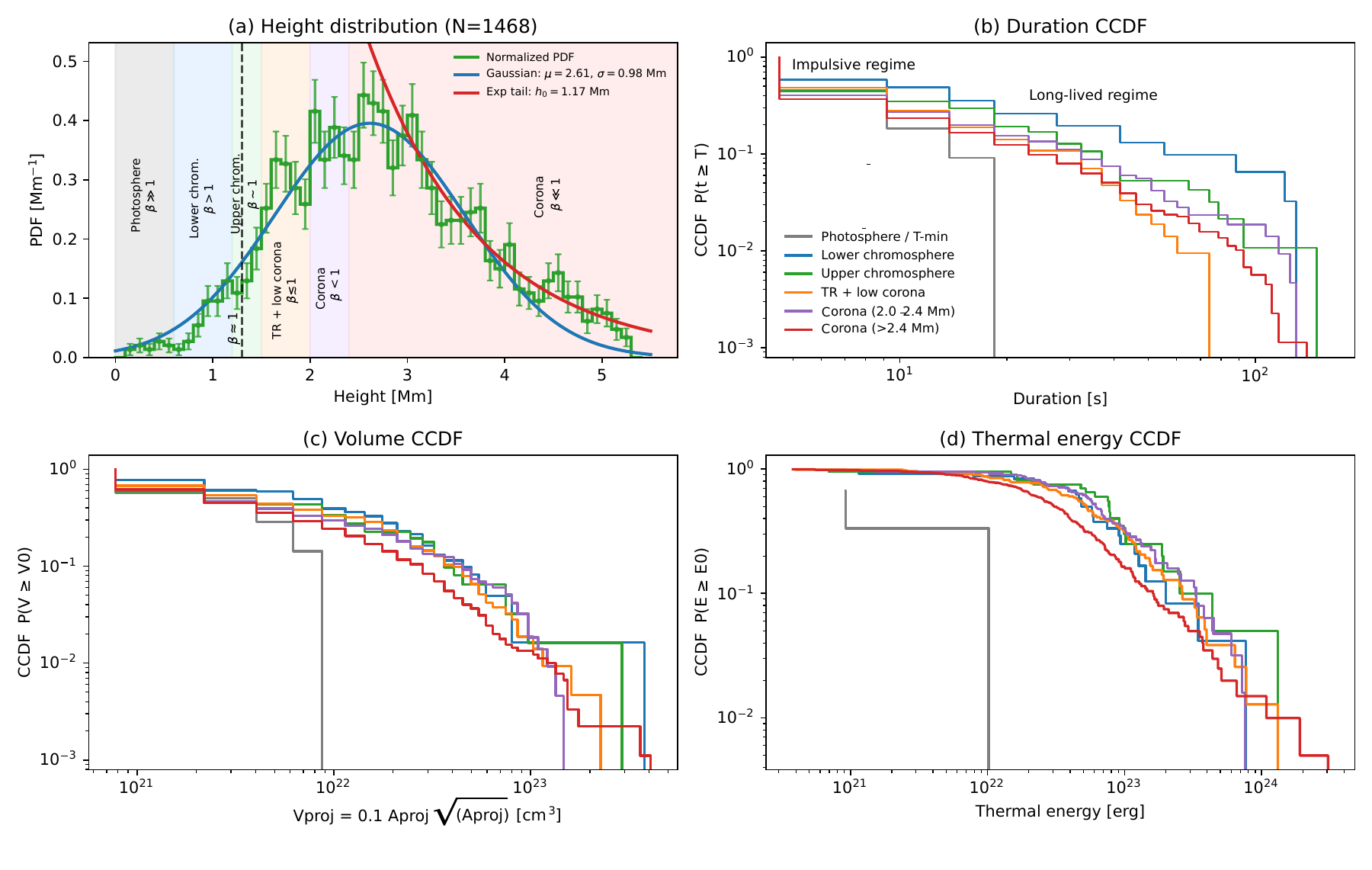}
\caption{Height dependence of campfire occurrence and properties following 
the quiet-Sun stratification of \citet{Judge2006}.
\textit{Panel (a)}:~Height PDF with atmospheric layers and plasma-$\beta$ 
regimes indicated. The core is well described by a Gaussian 
($\mu = 2.61$~Mm, $\sigma = 0.98$~Mm); the high-altitude excess 
is fitted by an exponential tail ($h_0 = 1.17$~Mm).
\textit{Panel (b)}:~Layer-resolved duration CCDFs (5~s cadence). Two 
regimes are visible: a short-duration impulsive regime common to 
all layers and a long-lived tail preferentially associated with 
chromospheric events. The tails extend to a few rare individual 
events of $\sim$150~s; the layer statistics in 
Table~\ref{tab:campfire_layer_summary} use the 90th and 95th 
percentiles, which are insensitive to these outliers.
\textit{Panel (c)}:~CCDFs of effective volume 
$V_{\rm proj} = 0.1\,A_{\rm proj}\sqrt{A_{\rm proj}}$.
\textit{Panel (d)}:~CCDFs of background-subtracted thermal energy. Unlike 
duration, volume and thermal-energy distributions are broadly similar 
across layers, indicating that height-dependent plasma conditions 
modulate dissipation timescales rather than the energy scale of 
impulsive heating.
}
\label{fig:composite_judge2006}
\end{figure*}

Each campfire was assigned to a quiet-Sun atmospheric layer following 
the height stratification of \citet{Judge2006}. 
Figure~\ref{fig:composite_judge2006} summarises the height dependence 
of key observables; Table~\ref{tab:campfire_layer_summary} lists 
layer-resolved statistics. All quantities in this section were derived using 
background-subtracted DEM diagnostics and the baseline model 
$V_{\rm proj} = 0.1\,A_{\rm proj}\sqrt{A_{\rm proj}}$.

The height distribution (Fig.~\ref{fig:composite_judge2006}a) peaks 
in the corona, around $\sim$2.6~Mm (Gaussian centre 
$\mu = 2.61$~Mm), and extends both downwards to chromospheric 
layers and upwards to greater coronal heights. The core follows a Gaussian; a high-altitude 
tail indicates that campfires span plasma-$\beta$ environments from 
$\beta \gtrsim 1$ at low heights to $\beta \ll 1$ in the corona.

The clearest height dependence is in event duration
(Fig.~\ref{fig:composite_judge2006}b). Short-duration events occur 
at all heights with comparable probability; the shortest detectable 
duration is set by the cadence ($t \lesssim 5$~s single frame, 
$t_{\min} \approx 10$~s for two frames). Beyond this floor, 
the lower chromosphere shows a clear excess of long-lived events 
extending to $\sim$80~s ($\sim$9\% of events there exceed 50~s, 
versus $\sim$3\% at greater heights), while coronal events are 
predominantly short-lived (Table~\ref{tab:campfire_layer_summary}). 
A rank correlation across all individual events, independent of any layer 
binning, confirms this trend at high significance 
(Spearman $\rho = -0.11$, $p \approx 2\times10^{-5}$): although 
modest in amplitude, the anti-correlation between height and 
duration is statistically robust and driven primarily by the 
chromospheric excess of long-lived events.

By contrast, emitting volumes and thermal energies show only weak 
height dependence: as seen in Fig.~\ref{fig:composite_judge2006}c,d, 
the layer-resolved CCDFs overlap closely and span comparable ranges, 
in contrast to the clear layer-to-layer separation seen for the 
duration distributions (Fig.~\ref{fig:composite_judge2006}b). The primary layer-to-layer variation is in event lifetime, not 
energy scale (Table~\ref{tab:campfire_layer_summary}).

These results indicate that height-dependent plasma conditions 
primarily modulate the timescales of campfire evolution. 
The presence of long-duration events at chromospheric heights, 
without a corresponding shift in volume or energy, suggests that 
comparable energy is released over similar spatial scales but 
dissipated more slowly in denser, higher-$\beta$ layers.

\subsection{Heating rates and radiative losses}
\label{sec:heating_vs_radiation}

\begin{figure}[t!]
\centering
\includegraphics[width=1.0\linewidth]{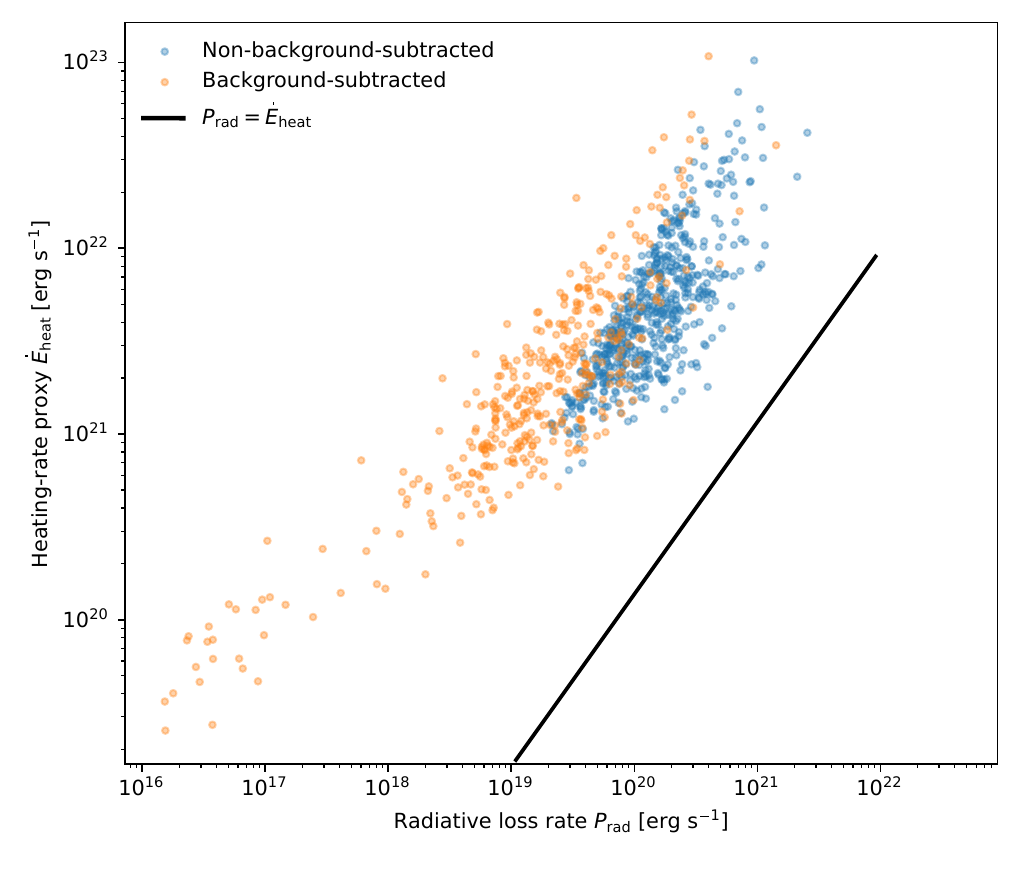}
\caption{Heating-rate proxy $\dot{E}_{\rm heat} = E_{\rm th}/t$ versus 
radiative loss rate $P_{\rm rad} = EM_{\rm vol}\,\Lambda(T)$, 
where $EM_{\rm vol} = EM_{\rm col}\,A_{\rm proj}$ and $\Lambda(T)$ 
is the optically thin radiative loss function. Blue and orange 
circles show non-background-subtracted and background-subtracted 
DEM values, restricted to events with robust thermodynamic solutions 
\citep{Berghmans2021}. The diagonal line marks 
$P_{\rm rad} = \dot{E}_{\rm heat}$. All events lie well above this 
line, indicating that radiative losses alone cannot balance the 
injected thermal energy; non-radiative losses -- primarily thermal 
conduction and enthalpy flux -- must therefore play an important 
role in the energy budget.
}
\label{fig:pheat_prad}
\end{figure}

To assess the energetic consistency of campfire detections, we 
compared a heating-rate proxy with the radiative loss rate for each 
event. We computed
\begin{equation}
EM_{\rm vol} = EM_{\rm col}\,A_{\rm proj},
\end{equation}
\begin{equation}
P_{\rm rad} = EM_{\rm vol}\,\Lambda(T),
\end{equation}
where $\Lambda(T)$ is the optically thin radiative loss function at 
the DEM-derived temperature, and defined the heating-rate proxy as
\begin{equation}
\dot{E}_{\rm heat} \equiv \frac{E_{\rm th}}{t},
\end{equation}
with $t$ the event duration. As shown in 
Fig.~\ref{fig:pheat_prad}, $\dot{E}_{\rm heat} \gg P_{\rm rad}$ 
for all events, confirming that radiative losses represent a 
negligible correction to the heating-rate estimate and that 
non-radiative cooling channels -- thermal conduction and enthalpy 
flux -- dominate the energy budget.

We estimated the total campfire contribution to quiet-Sun coronal heating by 
summing the event thermal energies and normalising by the observing time and 
field-of-view area:
\begin{equation}
F_{\rm heat} = \frac{\sum_i E_{{\rm th},i}}{A_{\rm FOV}\,t_{\rm obs}}
\quad\mathrm{[erg\,cm^{-2}\,s^{-1}]}.
\end{equation}
We obtain $F_{\rm heat} \sim 10^{2}$--$10^{2.5}$~erg~cm$^{-2}$~s$^{-1}$, 
a sub-percent fraction of the canonical quiet-Sun loss rate of 
$3\times10^5$~erg~cm$^{-2}$~s$^{-1}$ \citep{Withbroe1977}. This is 
a conservative lower limit: combining a lower $3\sigma$ detection 
threshold with the $\sim$26\% DEM completeness (389 of the 1,468 events 
with robust background-subtracted DEM solutions) could raise
the inferred contribution to a value of order a few percent of the
canonical requirement (an approximate, single-epoch estimate in which
the two corrections were treated as independent).

This result is consistent with earlier quiet-Sun studies. 
\citet{Ulyanov2019} find that directly detected events fall short 
of the required radiative losses by a factor of $\sim$30 and argue 
that the distribution must extend to lower energies to become 
sufficient. \citet{Krucker1998} reported that microflares account 
for $\sim$16\% of the quiet-corona output -- a higher fraction 
than ours, most likely reflecting their larger event energies and 
the different methodology -- their energies were derived from a 
two-filter line-ratio temperature estimate rather than the 
multi-channel DEM diagnostics used here. Similarly, 
\citet{Purkhart2022} find that SDO/AIA nanoflare-like 
events contribute $\sim$10\% of the quiet-Sun radiative losses. 
The lower fraction found for campfires reflects our more 
stringent detection threshold and shorter observing window 
rather than an intrinsically smaller energy contribution; 
with $\alpha > 2$, the energy budget is expected to be 
dominated by the smallest, currently undetected events.

A key theoretical result from \citet{Hudson1991} is that for 
$dN/dE \propto E^{-\alpha}$ with $\alpha > 2$, the total energy 
is dominated by the smallest events. With $\alpha \approx 2.2$, 
campfires could provide a pathway to coronal heating if the 
power law continues below current detection limits. We caution, 
however, that the observed distribution flattens below the fitted 
range, most likely owing to detection incompleteness rather than 
an intrinsic change in slope; whether the true slope remains above 
2 in this regime cannot be established from the present data and 
must be tested by more sensitive future observations. The sub-percent 
flux is therefore a lower bound; smaller, unresolved events could 
supply the missing energy. This is consistent with -- and does not 
contradict -- studies showing that individual campfires contain 
sufficient local magnetic energy to power themselves 
\citep[e.g.][]{Panesar2021}, which address the available energy 
reservoir rather than the area-averaged thermal-energy flux.

\subsection{Comparison with previous studies}
\label{sec:comparison}

\begin{figure}[t!]
\centering
\includegraphics[width=1.0\linewidth]{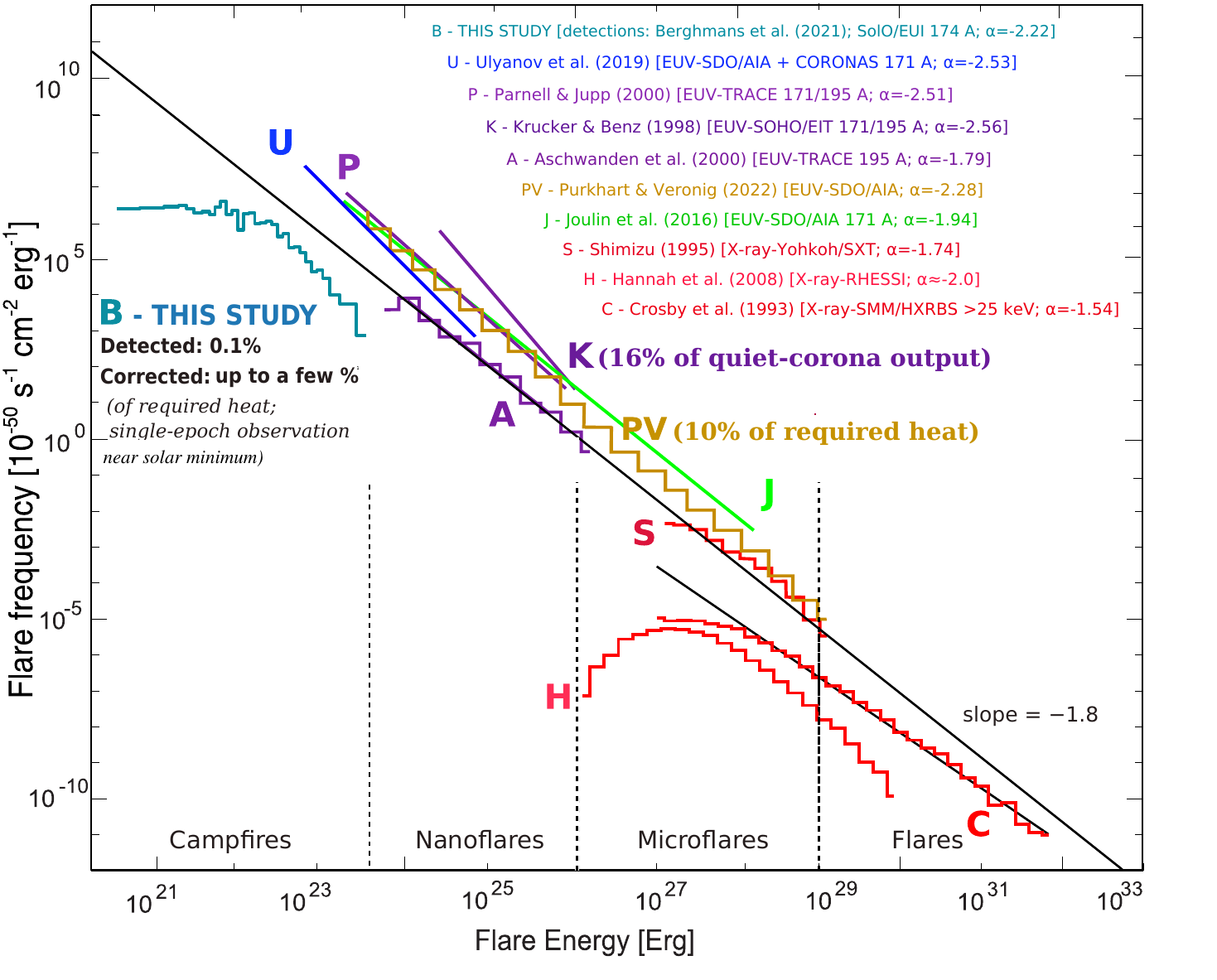}
\caption{Flare frequency distributions of campfire thermal energies compared
with EUV, soft X-ray (SXR), and hard X-ray (HXR) flare populations, following
\citet{Aschwanden2000}. The distribution is normalised per unit
time, area, and energy, allowing direct comparisons across event
classes. Campfires occupy the low-energy end of the flare hierarchy
with power-law slopes $\alpha = 1.99$--$2.38$ depending on the
volume model; the reference $A_{\mathrm{proj}}^{3/2}$-scaled model
gives $\alpha \approx 2.2$. The black diagonal line shows the
reference slope of $-1.8$ from \citet{Aschwanden2000}.
Power-law indices for previous studies are taken from the literature.
Curve~B (this study) is based
on the campfire detections of \citet{Berghmans2021}, with thermal
energies and the resulting frequency distribution derived here. It
lies below the EUV/X-ray studies because of our stringent
$\geq5\sigma$ detection threshold; a lower threshold would shift
it upwards. The heating-fraction percentages for the other studies
are their published values, shown for comparison.
}
\label{fig:eth_pdf_aschwanden}
\end{figure}

Figure~\ref{fig:eth_pdf_aschwanden} places campfires in the broader 
flare energy hierarchy, following the composite frequency distribution 
of \citet{Aschwanden2000}. Campfires occupy the low-energy extension 
of the flare distribution with power-law slopes $\alpha = 1.99$--$2.38$ 
depending on the volume model, consistent with the steep slopes 
required for small events to dominate the total energy budget 
\citep{Hudson1991}. Notably, campfires extend to energies below 
the lower limit of previous quiet-Sun nanoflare studies, probing 
the impulsive-heating population at the smallest scales yet 
observed.

Prior EUV and X-ray studies using the Solar and Heliospheric Observatory 
(SOHO), the Transition Region and Coronal Explorer (TRACE), CORONAS-F, SDO, 
Yohkoh, and the Reuven Ramaty High Energy Solar Spectroscopic Imager (RHESSI) 
identified nanoflares and microflares
with thermal energies 
spanning $\sim10^{23}$--$10^{26}$~erg, depending on temperature 
coverage, background treatment, and geometric assumptions 
\citep{Crosby1993, Shimizu1995, Krucker1997, Benz1998, 
Parnell2000, Hannah2008}. The campfires analysed here populate the 
low-energy extension of this hierarchy, with background-subtracted 
thermal energies spanning $\sim10^{20}$--$10^{24}$~erg. The 
statistically robust power-law regime extends over one to two 
decades depending on the volume model: from $\sim10^{22}$~erg for 
$A_{\mathrm{proj}}^{3/2}$-scaled models to $\sim10^{23}$~erg for 
the Aschwanden model, up to $\sim5\times10^{24}$~erg in all cases. 
Non-background-subtracted distributions shift upwards by $\sim$one 
order of magnitude, consistent with expectations for line-of-sight 
integrated EUV emission.

We note that below the fitted regime, in the range 
$\sim10^{20}$--$4\times10^{22}$~erg, the distribution becomes 
flatter than the power-law tail. This flattening at the low-energy 
end is most likely caused by incomplete detection -- a combination 
of the $\geq5\sigma$ detection threshold, the limited sensitivity 
of \hrieuv, and the finite spatial resolution, which together cause 
the faintest and smallest events to be missed. We therefore fitted the 
power law only above $E_{\min}$, where the data are statistically 
robust, and did not treat the flat low-energy part as a real feature. 
As a result, our estimate of the campfire energy contribution is 
conservative: if the true distribution continues with $\alpha > 2$ 
below the completeness limit, the energy in smaller events would 
exceed what the observed flattening suggests. Probing this regime 
directly will require observations at closer perihelia, where the 
improved resolution and sensitivity of Solar Orbiter/EUI can push 
the completeness limit to lower energies. This energy range is supported by 
3D MHD simulations: \citet{Chen2021} report thermal energies of 
$10^{22}$--$10^{24}$~erg for simulated campfire-like events, 
providing an important cross-validation of both the observational 
estimates and the adopted thermodynamic and geometric assumptions.

\begin{figure}[t!]
\centering
\includegraphics[width=1.0\linewidth]{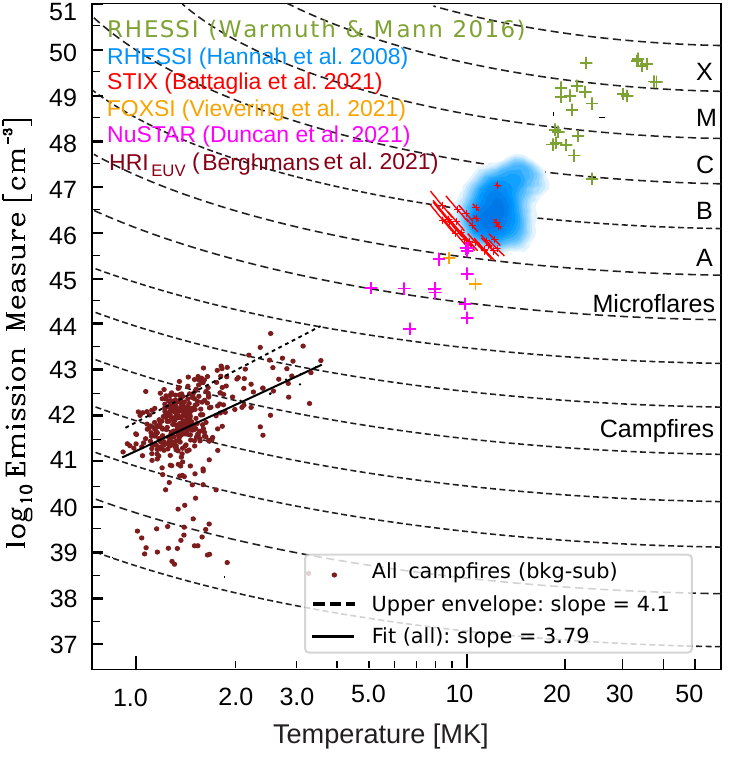}
\caption{$EM$--$T$ diagram comparing EUI/HRI$_{\rm EUV}$ campfires 
(dark red circles; background-subtracted 
$EM_{\rm vol} = EM_{\rm col}\,A_{\rm proj}$) with larger solar 
flares from RHESSI \citep{Hannah2008, Warmuth2016A&A}, FOXSI 
\citep{Vievering2021}, \textit{NuSTAR} \citep{Duncan2021}, and 
Solar Orbiter/STIX \citep{Battaglia2021}. The blue density map 
shows the RHESSI microflare population \citep{Hannah2008}. 
Campfire $EM$ and $T$ were first reported by 
\citet{Berghmans2021}; here we re-analysed the same events with 
background subtraction. Both campfire fits use the 
background-subtracted data: the solid line shows a power-law fit 
to all campfire events ($EM_{\rm vol} \propto T^{3.79}$), while 
the straight dashed line shows the fit to their upper envelope 
(slope~$= 4.1$), consistent with the $EM \propto T^{4}$ scaling 
of larger flares \citep{Warmuth2016A&A}. The faint curved dashed 
lines spanning the diagram mark contours of constant GOES soft 
X-ray flux, with the corresponding GOES classes indicated on the 
right axis. Campfires extend the $EM$--$T$ continuum towards the 
lowest energies, connecting smoothly to the microflare 
population. This figure extends the $\mathit{EM}$--$T$ comparison 
of \citet{Battaglia2021} to include EUI campfires at the 
low-energy end.
}
\label{fig:eui_stix}
\end{figure}

Figure~\ref{fig:eui_stix} places campfires in the $EM$--$T$ context 
of Solar Orbiter/STIX  microflares. For the background-subtracted campfire data, the fit to all events 
($EM_{\mathrm{vol}} \propto T^{3.79}$) and the upper-envelope fit 
(slope~$= 4.1$) are consistent both with the background-subtracted 
upper-envelope scaling derived independently in 
Fig.~\ref{fig:emt_scaling} (slope~$= 4.1$) and with relations 
reported for larger flares \citep{Warmuth2016A&A}.
Although STIX microflares probe significantly hotter plasma, their 
$EM$--$T$ scaling follows the same qualitative trends as campfires, 
shifted towards higher temperatures and emission measures, indicating 
a common thermodynamic origin. Campfires thus extend the $EM$--$T$ 
continuum of impulsive energy release from microflares down to 
picoflare energies, bridging the two populations without a 
discernible gap. Together with the layer-resolved analysis, this 
supports the view that campfires, nanoflares, and microflares form 
a continuous hierarchy distinguished by their atmospheric environment 
and cooling timescales -- where event duration serves as a proxy 
for the EUV cooling time -- rather than by fundamentally different 
energy-release mechanisms.

Following \citet{Ulyanov2019}, we estimated how far below the 
detection limit the power law would need to extend for the 
integrated event energy to match the canonical heating flux 
of $3\times10^5$~erg~cm$^{-2}$~s$^{-1}$ \citep{Withbroe1977}. 
Integrating the power-law distribution down to a lower energy 
$E_{\mathrm{low}}$ and equating the result to the required flux 
yields $E_{\mathrm{low}} \approx 2.4\times10^{17}$~erg for 
$\alpha = 2.2$, rising to $\approx10^{21}$~erg for $\alpha = 2.5$. 
This strong sensitivity to $\alpha$ reflects the steepness of 
the energy integral: steeper slopes accumulate the required 
energy over a narrower range and therefore need not extend 
to such low energies.

\citet{KatsukawaTsuneta2001} provide a useful order-of-magnitude 
benchmark: meeting the heating requirement with $10^{20}$~erg 
nanoflares demands $10^6$~s$^{-1}$ in a single active region -- 
roughly six orders of magnitude more frequently than our detected 
campfires. This gap reflects our more stringent $\geq5\sigma$ 
threshold; at $3\sigma$ the inferred contribution already reaches 
$\sim$1.2\%. Accounting further for the $\sim$26\% DEM completeness 
could add a further factor of a few, bringing the total to a value of 
order a few percent overall.
Additionally, our energies capture only the coronal ($\sim$1~MK) 
component; campfire plasma also emits in cooler transition-region 
lines \citep{Dolliou2023, Dolliou2024, Huang2023}, so the true 
energy budget extends beyond EUV imaging alone. Coordinated EUI 
and SPICE observations are uniquely placed to constrain the full 
campfire energy budget \citep{Narang2025}. As a remote-sensing 
estimate from a single epoch, the precise fraction cannot be 
pinned down definitively; it depends on the observing region, the 
phase of the solar cycle, and the detection limits. We therefore 
present it as a robust order-of-magnitude result rather than a 
final figure. Its significance lies less in the exact percentage 
than in the fact that an impulsive-heating population is now detectable 
down to picoflare energies and is beginning to be resolved by atmospheric 
layer.

\subsection{Limitations of our analysis}
\label{sec:limitations}

The thermal energies represent only the excess thermal component 
from EUV emission, not the total available magnetic energy. Other 
energy channels -- bulk motions, waves, non-thermal particles 
\citep{warmuth2020thermal} -- are not constrained here, so the 
inferred values are lower limits to the total energy released.

Our sample may also be incomplete in terms of temperature. The 
\hrieuv\ 174~\AA\ passband responds mainly to plasma near 
$\sim$1~MK, so events that heat plasma predominantly to higher or 
lower temperatures would emit weakly in this channel and could 
escape detection entirely. Such events would be absent from our 
catalogue rather than merely underestimated in energy, so the 
detected population, and hence the inferred heating contribution, 
represents a lower limit also in this respect.

The dominant systematic uncertainty is the emitting volume, which 
cannot be measured directly. The four geometric models explored here 
span the range commonly used in the field; their consistency 
demonstrates that the main statistical conclusions are robust to 
geometric assumptions.

Temperatures and emission measures come from the DEM analysis of 
Paper~I \citep{Berghmans2021}, performed using the regularised 
inversion method of \citet{Hannah2012}. Different DEM inversion methods can 
yield somewhat different temperature estimates \citep{Hofmeister2025}; 
the uncertainties in Paper~I reflect internal consistency rather 
than the full systematic range. Since
\[
E_{\mathrm{th}} \propto T\,\sqrt{\mathrm{EM}_{\mathrm{vol}}\, V},
\]
fractional emission-measure uncertainties propagate as half their 
value into $E_{\mathrm{th}}$, while fractional temperature 
uncertainties propagate directly. Emission measure is therefore the dominant 
source of error for individual events, though population-level 
distributions -- shape, slope, dynamic range -- are unaffected.

Finally, these results derive from a single 245~s sequence near 
solar minimum, ensuring a clean quiet-Sun sample but limiting 
generalisability to other activity levels or magnetic environments. 
Future multi-epoch observations will be needed to assess how 
campfire properties vary with solar cycle phase.

\section{Discussion}

Individual current sheets do not dissipate in isolation 
\citep{Vlahos1994}; the observed EUV response reflects the collective 
thermal effect of many unresolved dissipation sites distributed along 
the magnetic structure. This picture is supported by 3D MHD simulations 
that reproduce campfire EUV brightenings as the plasma response to 
localised current dissipation, without resolving the underlying 
kinetic-scale sheets \citep{Chen2021}. EUV campfires therefore 
constrain the macroscopic plasma response -- spatial extent, 
characteristic timescales, and thermal signatures -- rather than 
the dissipation process itself.

The height dependence of event duration 
(Fig.~\ref{fig:composite_judge2006}) -- driven by the excess of 
long-lived events in the lower chromosphere -- without an 
analogous dependence in energy or volume, provides a clear observational anchor for 
height-dependent dissipation and cooling regimes. Perpendicular 
current sheets dissipating via magnetic reconnection are expected to 
reach kinetic scales of tens to hundreds of metres. By contrast, in 
the low corona where $\beta \sim 1$, quasi-parallel current sheets 
are predicted to be systematically wider -- by roughly an order of 
magnitude -- owing to lower dissipation thresholds and reduced 
magnetic shear \citep{Demoulin2000}. Such broader structures favour 
gentler, thermally dominated energy release on sound-speed-limited 
timescales of tens to hundreds of seconds, consistent with the 
observed excess of long-duration campfires at chromospheric heights 
\citep[e.g.][]{Pontin2017}. The observed duration distribution, 
spanning both rapid ($\sim$10~s) and gradual ($\sim$100~s) events, 
may thus reflect a transition from fast Alfv\'enic dissipation in 
low-$\beta$ coronal plasma to slower, thermally dominated dissipation 
in higher-$\beta$ plasma near the transition region. The absence of 
a height dependence in thermal energy indicates that comparable 
amounts of magnetic energy are released over similar spatial scales 
but on different timescales, consistent with both perpendicular 
reconnection and parallel current dissipation operating in the low 
corona \citep{Klimchuk2015,Klimchuk2017}. Taken together, these results 
reinforce the view that EUV observations capture thermodynamic envelopes of 
unresolved, intermittent magnetic energy dissipation -- precisely as 
anticipated by modern coronal-heating theory.

\section{Conclusions}

We have presented a thermal-energy analysis of EUV campfires observed 
in the quiet Sun by Solar Orbiter/EUI \hrieuv\ and catalogued 
by \citet{Berghmans2021}, building on the thermodynamic diagnostics of 
Paper~I. We quantified thermal energies, heating-rate proxies, and 
statistical distributions using multiple geometric volume models, and 
compared the campfire population with Solar Orbiter/STIX microflares 
\citep{Battaglia2021}, SDO/AIA nanoflare-like events \citep{Purkhart2022}, 
and earlier flare statistics. Our main conclusions are:

\begin{enumerate}

\item Campfires occupy the picoflare energy regime, with 
background-subtracted thermal energies spanning 
$\sim10^{20}$--$10^{24}$~erg and rare events reaching 
$\sim5\times10^{24}$~erg. Their high-energy tails follow power-law 
behaviour over $\sim$1.5--2 decades with slopes 
$\alpha \approx 2.0$--$2.4$ depending on the adopted volume model; 
the reference $A_{\mathrm{proj}}^{3/2}$-scaled model gives 
$\alpha \approx 2.2$, consistent with a low-energy extension of the 
solar flare hierarchy.

\item The $\mathit{EM}$--$T$ scaling of campfires 
($\mathit{EM}_{\mathrm{vol}} \propto T^{3.79}$, upper-envelope 
slope~$= 4.1$) is consistent with relations reported for larger 
flares \citep{Warmuth2016A&A}, supporting their interpretation as 
the low-energy extension of impulsive heating in magnetically 
confined plasma and connecting smoothly to the microflare population 
without a discernible gap.

\item The statistical properties -- energy range, distribution shape, 
and power-law slope -- are robust against reasonable variations in 
source geometry, background subtraction, and instrumental 
uncertainties, indicating that the observed scaling is intrinsic to 
the campfire population.

\item Campfires span atmospheric heights from the chromosphere into 
the corona, with the occurrence distribution peaking around 
$\sim$2.6~Mm, covering plasma-$\beta$ regimes from 
$\beta \gtrsim 1$ to $\beta \ll 1$, with no evidence of discrete 
sub-populations.

\item Campfire durations exhibit a clear height dependence: short 
events occur at all heights, but the lower chromosphere shows an 
enhanced fraction of long-lived brightenings ($\sim$9\% exceed 
50~s, versus $\sim$3\% at greater heights). Emitting volumes and 
thermal energies, by contrast, are broadly similar across layers 
(with at most a slight decrease in the high corona). This 
indicates that height-dependent plasma conditions modulate dissipation 
and cooling timescales rather than the characteristic energy scale -- 
consistent with slower current dissipation in higher-$\beta$ plasma 
near the transition region.

\item Heating-rate proxies exceed radiative loss rates at all heights, 
confirming that campfires are heating-dominated impulsive events. 
Directly detected campfires contribute $\sim$0.1\% of the canonical 
quiet-Sun heating requirement \citep{Withbroe1977}. This is a 
conservative lower limit: lowering the detection threshold to 
$3\sigma$ and accounting for the $\sim$26\% DEM completeness could 
together raise the contribution to a value of order a few percent 
(up to $\sim$5\%) of the canonical requirement. This combined 
factor is an order-of-magnitude estimate: the two corrections 
were treated as independent, and the underlying Poisson and threshold 
uncertainties were not propagated in detail. With $\alpha > 2$ 
\citep{Hudson1991}, the energy integral is dominated by the smallest 
events; if the power law extends with the same slope below our 
detection limit, unresolved events could supply a significant 
additional fraction. Whether the slope remains this steep in the 
incompletely sampled low-energy range remains to be established by 
more sensitive observations. Coordinated 
EUI and SPICE observations, combined with lower detection thresholds, 
are therefore the most direct path towards a complete census of the 
campfire energy budget.

\item These results are based on a single observing sequence near 
solar minimum, providing a clean quiet-Sun sample. Multi-epoch 
observations across the solar cycle will be needed to establish 
whether the height-resolved trends reported here are universal.

\end{enumerate}

The overarching finding of this work is that EUV campfires extend 
the impulsive-heating population to the lowest energies yet probed 
in the quiet Sun, although their full energetic role remains to be 
quantified. The directly detected population accounts for a 
measurable fraction of the heating requirement ($\sim$0.1\%). 
Accounting approximately for the detection threshold and DEM 
completeness raises this estimate to a value of order a few percent, though we 
stress that this is a single-epoch, order-of-magnitude estimate 
without full uncertainty propagation. Whether the contribution 
increases further at still lower energies cannot be determined from the present data; 
the steep power-law slope ($\alpha > 2$) means this remains an open 
possibility, to be tested by future observations reaching below 
our current detection limit.

It is important to distinguish two complementary aspects of the 
campfire energy budget. Studies of individual campfires have shown 
that their available magnetic energy is sufficient to power the local 
brightening \citep{Panesar2021} -- establishing that the energy source 
exists at the level of individual events. The present work addresses 
a different and harder question: whether the integrated thermal 
output of the campfire population, summed over the field of view and 
normalised to the quiet-Sun area, can account for the global coronal 
heating requirement. Our results show that the detected population 
falls short. The associated uncertainties are large enough that 
we cannot rule out a substantial contribution from events below 
the current $\geq5\sigma$ detection threshold: with $\alpha > 2$, 
such unresolved events could in principle supply a significant 
additional fraction of the heating requirement. Two conditions must 
be simultaneously satisfied for campfire-like events to dominate 
coronal heating \citep{Hudson1991, KatsukawaTsuneta2001}: the 
energy distribution must follow a power law with $\alpha > 2$, 
so that the energy integral diverges towards smaller events; and 
the occurrence rate of small events must be sufficiently high. 
Our measured slope of $\alpha \approx 2.2$ satisfies the first 
condition. The second remains to be tested observationally at 
energies below our current detection threshold -- precisely the 
regime accessible to future Solar Orbiter campaigns at 
closer perihelia and with coordinated SPICE spectroscopy.

The key physical insight from the layer-resolved analysis is that 
atmospheric stratification primarily regulates dissipation and thermal 
relaxation timescales rather than the energy scale of impulsive 
heating. This provides a new observational constraint distinguishing 
between current-dissipation models operating in different plasma-$\beta$ 
regimes. Future coordinated observations with Solar Orbiter 
EUI and SPICE, at progressively closer perihelia and with lower 
detection thresholds, offer the most direct route to resolving whether 
the campfire population -- and the broader class of small-scale EUV 
brightenings -- can account for a dominant fraction of the energy 
required to maintain the quiet-Sun corona.

\begin{samepage}
\section*{Code availability}
A Python implementation of the methods for calculating campfire volumes,
emission measures, and thermal energies is available at
\url{https://github.com/epodlad/campfire-thermal-energy}.
The repository was released after journal publication and documents the
formulas, assumptions, input requirements, and validation status.\par
\end{samepage}

\begin{acknowledgements}
We thank the anonymous referee for constructive comments that helped 
clarify the interpretation of the results and motivated a more coherent 
synthesis of impulsive heating, current dissipation, and plasma 
processes in the low solar corona.

The Solar Orbiter mission is a space mission of international 
collaboration between ESA and NASA, operated by ESA. The EUI instrument 
was built by a consortium led by CSL, including IAS, MPS, MSSL/UCL, 
PMOD/WRC, ROB, and LCF/IO, with funding from the Belgian Federal 
Science Policy Office (BELSPO/PRODEX PEA 4000134088, 4000112292, and 
4000106864); the Centre National d'Etudes Spatiales (CNES); the UK 
Space Agency (UKSA); the Bundesministerium f\"ur Wirtschaft und Energie 
(BMWi) through the Deutsches Zentrum f\"ur Luft- und Raumfahrt (DLR); 
and the Swiss Space Office (SSO).

The ROB team thanks BELSPO for support through the ESA-PRODEX programme 
(grants 4000134088, 4000112292, 4000136424, and 4000134474). 
D.M.L.\ is grateful to the UK Science and Technology Facilities Council 
(STFC) for an Ernest Rutherford Fellowship (ST/R003246/1). 
P.A.\ acknowledges funding from an STFC Ernest Rutherford Fellowship 
(ST/R004285/2). 
O.P.\ is grateful to the German Leibniz Association and the MWFK 
Brandenburg. 
The work of F.S.\ was supported by DLR grant 50~OT~1904. 
A.M.V.\ and S.P.\ acknowledge the Austrian Science Fund (FWF): 
project 10.55776/I4555-N.
\end{acknowledgements}

\end{document}